\documentclass[aps,twocolumn]{revtex4}
\usepackage{epsfig}
\usepackage{psfrag}
\usepackage{color}
\newcommand{\be}{\begin{equation}}
\newcommand{\ee}{\end{equation}}
\newcommand{\bea}{\begin{eqnarray}}
\newcommand{\eea}{\end{eqnarray}}
\newcommand{\ba}{\begin{array}}
\newcommand{\ea}{\end{array}}

\usepackage{natbib}

\begin{document}

\title{Aftermath epidemics: Percolation on the sites visited by generalized random walks}

\author{Mohadeseh Feshanjerdi}\email{(corresponding author) m.feshanjerdi@alzahra.ac.ir}
\affiliation{Department of Condensed Matter Physics, Faculty of Physics, Alzahra University, P. O. Box 1993893973, Tehran, Iran}
\author{Amir Ali Masoudi}\email{(corresponding author) masoudi@alzahra.ac.ir}
\affiliation{Department of Condensed Matter Physics, Faculty of Physics, Alzahra University, P. O. Box 1993893973, Tehran, Iran}
\author{Peter Grassberger}\email{p.grassberger@fz-juelich.de}
\affiliation{JSC, FZ J$\ddot{u}$lich, D-52425 J$\ddot{u}$lich, Germany}
\author{Mahdieh Ebrahimi}\email{evrusebr@gmail.com}
\affiliation{Institute of Condensed Matter Physics, Technical University of Darmstadt, Hochschulstr. 6, 64289 Darmstadt, Germany}

\date{\today}

\begin{abstract}
We study percolation on the sites of a finite lattice visited by a generalized random walk of finite 
length with periodic boundary conditions. More precisely, consider Levy flights and walks with finite 
jumps of length $>1$ (like Knight's move random walks (RWs) in two dimensions and generalized Knight's 
move RWs in 3D). In these walks, the 
visited sites do not form (as in ordinary RWs) a single connected cluster, and thus percolation 
on them is nontrivial. The model essentially mimics the spreading of an epidemic in a population 
weakened by the passage of some devastating agent -- like diseases in the wake of a passing army or 
of a hurricane. Using the density of visited sites (or the number of steps in the walk) as a 
control parameter, we find a true continuous percolation transition in all cases except for the 2D 
Knight's move RWs and Levy flights with Levy parameter $\sigma \geq 2$. For 3D generalized Knight's 
move RWs, the model is in the universality class of pacman percolation, and all critical exponents 
seem to be simple rationals, in particular $\beta=1$. For 2D Levy flights  with $0 <\sigma < 2$, scale
invariance is broken even at the critical point, which leads at least to very large corrections in 
finite-size scaling, and even very large simulations were unable to unambiguously determine  the 
critical exponents.

\end{abstract}

\maketitle

\section{Introduction}

One disaster often does not come alone. In the present paper we deal with the purely geometric
-- i.e., percolation -- aspects of an epidemic which comes in the wake of another disaster like a 
hurricane or a war \cite{Smallman}, and can spread only on the sites weakened by the first.

Percolation in its simplest version (called OP in the following) deals with the establishment of 
long range connectivity in random but statistically homogeneous systems with only short range 
links between its units \cite{Stauffer,Sahimi}. The two best known examples of OP are site and 
bond percolation, where the system is a regular lattice of finite dimension, and local links 
are established by inserting sites or bonds \cite{Stauffer}. 

This is one of the paradigmatic models in statistical physics and has many applications, the 
most important one being the spreading of epidemics \cite{Bailey}. Starting from a local seed, a system-wide 
epidemic (or pandemic) can evolve only, if the spreading agent (virus, bacterium, or even rumor)
can reach wide regions, i.e., if large clusters of sites are connected. If the population is 
originally healthy and susceptible (except for the seed), and becomes immune or dead after
a finite time of illness, this is the so-called SIR (susceptible-infected-removed) model
\cite{Kermack,Grassberger1983}.

There are of course many modifications of this simple scenario \cite{Saberi,Araujo}: 

-- The system is not a regular lattice, but some sort of network  \cite{Newman,referee2-1,referee2-2}.
This leads to new universality classes, but at least if the network is close to regular (all nodes have  
similar degree) and uncorrelated, the situation is similar as to a regular lattice.

-- When recovered individuals become susceptible again, the resulting SIS model is in a 
different universality class from SIR or OP \cite{Hinrichsen}.

-- If there are finite incubation or latency periods between exposure to the 
spreading agent and the development of symptoms, in the resulting SEIR (susceptible-exposed-infectious-removed) 
compartmental model \cite{Bailey,referee1-1} the universality class is in general not changed.

-- Things change again, if contact with more than one infectious neighbor is needed to 
infect a susceptible individual. In the extreme case of bootstrap and $k$-core percolation
\cite{Adler,di_Muro}, clusters can grow (or do not shrink) only, if new (old) sites have a certain
minimal number of neighbors in the cluster. This can be relaxed so  infection of a new 
site is more likely if it has more infected neighbors \cite{Janssen, Bizhani,Dorogovtsev2006}. 
The most dramatic effect in such cases is that the percolation transition can become
discontinuous or, actually, hybrid: Although the order parameter jumps at the transition 
point, one also observes  scaling laws as for continuous transitions.

-- Similar cooperativity effects occur, if two (or more) diseases cooperate in the sense that 
infection by one also makes the individual more susceptible to be infected by the other
\cite{Cai}.

-- Very important, in particular in modern times where people can carry infections over very
long distances by flights, are nonlocal single links. Often, this is modeled by assuming 
that the infectious agent can perform a Levy flight, i.e., the probability for a link between
two sites is described by a power law \cite{Grassberger1986,Janssen,Linder,Grassberger2012,Grassberger2013,Gori}.
In this case one finds continuous transitions in new universality classes which depend on
the value of the power-law exponent.

-- While long-range effects are treated in the above models as long-range 
contacts between static individuals, more realistic models take into account that 
individuals can move \cite{referee1-2, Belik}. In this case 
the connection with percolation is, strictly spoken, lost, because there is no static infected cluster
when the epidemic has ended. If the movements are slow, this may not be a big problem and 
the standard scaling laws could still hold with minor adaptions, but in case of Levy flights all 
scaling laws have to be re-considered \cite{referee1-2, Belik}.

-- In OP, new local connections are established randomly. In contrast, in explosive
percolation \cite{Achlioptas, Christensen} (EP) one inserts new connections such that the occurrence of 
large clusters is delayed. The percolation transitions in EP were first thought to be 
discontinuous, but they are actually continuous. Apart from the smallness of the order 
parameter exponent $\beta$, its most striking feature is that
for finite lattice sizes $L$, the width of the critical region and its shift relative 
to the infinite lattice critical point satisfy power laws with different exponents
\cite{Bizhani} -- at least when analyzed in the conventional way where the transition 
point is defined as independent of the individual realization of the process 
\cite{Li2023}.

-- The system can be non-homogeneous in the sense that some regions are more susceptible 
and others less so. This can lead to multiple percolation transitions, such that
changes in cluster size are of order $N$ in each, where $N$ is the size of the system
\cite{Bianconi}.

-- Even if the system is homogeneous on large scales, it might be that there are long-range 
correlations between the densities of susceptible individuals and/or the links. This 
is called correlated percolation (CP), and is maybe the largest and most varied class of 
nontrivial percolation models \cite{Correlated Percolation referee2}. 

It is this class of models which is considered in the present paper.

By far the best studied special case
is the Ising model. It is well known that the Ising critical point can be understood as 
a percolation type transition for carefully defined (Fortuin-Casteleyn) clusters \cite{Fortuin}. But one 
can also study the percolation of clusters defined simply as connected sets of + and - spins,
and of the boundaries between them. This was recently done by Grady \cite{Grady}, who
found in three dimensions a true percolation transition which is not in the OP universality
class. Remarkably, Grady found that, as in EP, the width of the 
critical region for finite $L$ and its shift from the exact critical point at $L=\infty$
scale with different exponents.

Another class of CP models is one where the correlations are assumed to decay with 
power laws $C(r) \sim r^{-\alpha}$, without specifying the mechanism which generated them 
\cite{Weinrib-Halperin,Weinrib,Schrenk}. Whether the resulting percolation transition is in the OP class
or not should now depend on $\alpha$, according to a generalized Harris criterion: 
The universality class should be modified iff $d\nu^{(0)} > \alpha$, where $\nu^{(0)}$ is 
the correlation length exponent of the model without long-range correlations. This 
is seen in Ref. \cite{Schrenk} for some critical exponents, but not for all.

Finally, there is pacman percolation \cite{Abete,Kantor}. In this case,
all sites are susceptible initially. But before the actual percolation process starts,
a random walker performs a walk (with periodic boundary conditions) of $T$ steps, where
$T \sim N$, with $N = L^d$ being the number of sites. Percolation is then considered 
only on those sites which were {\it not} visited by the walker. 

The model studied in the present paper can be seen as the opposite of pacman percolation:
We again have a finite-time random walk (RW) before the percolation proper takes place, but 
now the percolation process can take place only on sites that {\it had} been visited by
the walker. A real-world scenario which might be modeled by this is an army or a hurricane 
that passes through some geographic region, and an epidemic which can evolve only in the 
areas devastated by them. It is true that hurricanes in the Caribbean don't 
make RWs, but Timur's armies in Iran and neighboring countries \cite{Timur} and the 
armies in the Thirty years War in Germany \cite{Germany} came very close. Periodic 
boundary conditions are used both for the walker and for the percolation process.

An immediate problem with such a type of models is that the visited sites are connected 
for an ordinary RW, and thus the problem of percolation seems trivial. The
way out of this dilemma is, of course, to modify the walk such that visited sites are not 
(necessarily) connected. In the present paper, we study two such modifications:

(a) {\it Knight's move and next-nearest neighbor (NNN) move RW}. A Knight's move in chess is one 
where one moves two lattice constants in one direction (say, x), and one in the other (say, y). 
From a given position, there are eight such moves. A NNN move RWs (NNN-RW) is a walk
where one moves $\pm 1$ step in each direction. In the following, we shall only show results
for the Knight's move RW, but we have also done  extensive simulations for NNN-RWs. We will show 
that there is no sharp
percolation transition in this model in two dimensions, but there is one if the model is 
generalized to 3D. A Knight's move in this generalized 3D walk is one where one moves two
lattice constants in one direction and one in each of the two others. In this case there
are 24 moves.

(b) {\it Levy flights}. Here, the probability for a step to have a length $>r$ decreases for large 
$r$ as 
\be
    P(r) \sim r^{-\sigma}
\ee
with $0 < \sigma < 2$. Here, we studied only two dimensional lattices.
For $\sigma \to 0$, the walk is just a sequence of random jumps, 
and our model reduces to site percolation. For $\sigma >2$, the walk is in most respects
equivalent to a RW, except that visited sites do not necessarily form a single 
connected cluster. It is for the latter reason that we also studied  the case $\sigma=2.5$, 
to verify that the behavior is the same as for the Knight's move RW. We also studied the case 
$\sigma=2$, which is at the border between Levy flights and ordinary walks.

A particular feature of the present model is that the finite value of $T$ can induce, for 
finite $L$, an additional characteristic length scale. For RWs, this length scale would be the 
square root of the rms end-to-end distance 
\be 
   \sqrt{\langle R^2} \rangle \sim T^{1/2} \sim L^{d/2}, 
\ee
which diverges for $d > 2$ faster than $L$ when $L\to\infty$, if the periodic boundary conditions would not
bring it down to $L$. Indeed, as shown in Ref. \cite{Kantor}, the latter implies that the correlation 
between visited sites decays as $C(r) \sim r^{2-d}$. For Levy flights, different powers of $R$ scale differently, 
$\langle R^q \rangle \sim T^{q/\sigma}$, if $q>\sigma$ \cite{Mandelbrot}, and the correlation 
function $C(r)$ is, in general, not a power law (see Appendix). Thus it is not scale-free,
suggesting that several new length scales might be involved. This might imply that the standard 
finite size scaling (FSS) behavior is no longer valid for Levy flights, and that, in particular, 
the width and the shift of the critical peak in variables like the fluctuations of the 
order parameter might scale with different exponents, as found also in EP \cite{Christensen}
and in boundary percolation in the Ising model \cite{Grady}.

\section{Definitions of the models, algorithms, and computational details}

Both models live on square, respectively, cubic lattices. For computational efficiency, we 
replaced the periodic boundary conditions by helical ones, where one uses a single integer to 
label sites and neighbors of site $i$ are $(i\pm 1)\; {\rm mod}\; N, (i\pm L)\; {\rm mod}\; N, \ldots
(i\pm L^d) \;{\rm mod}\; N$.  For generating Levy flights, we used the 
algorithm of  Refs. \cite{Linder, Grassberger2013}: 
First, two random numbers ($\delta x, \delta y$) between 0 and 1 are chosen randomly. 
If $r^2 \equiv \delta x^2 + \delta y^2 \geq1$, they are discarded and a new pair is chosen. 
OtheRWsise, $\Delta x = \pm \frac{\delta x}{r^{1+2/\sigma}}$ and 
$\Delta y = \pm \frac{\delta y}{r^{1+2/\sigma}}$, where all four sign combinations are 
chosen with equal probability.

In the following, we shall use the words walk and walker both for Levy walks and 
for (generalized) Knight's move walks.

In the Introduction, walk and percolation were discussed as independent and subsequent 
parts of the model, but for computational efficiency we measured the properties 
related to percolation already during the walk by means of the site insertion version 
of the Newman-Ziff (NZ) algorithm \cite{Newman-Ziff}. In our algorithm, we keep track of 
the number $n$ of sites visited by the walker (we use $\rho =n/N$ as control parameter)
and the size $S_n$ of the largest cluster when $n$ sites are visited ($S_n/N$ is used as 
order parameter). At each step of the 
walk we registered whether a new site was visited or not. In the latter case, the next 
step was taken immediately. If a new site $i$ was visited, however, we increased the number 
$n$ of visited sites by 1 and performed one step of the NZ algorithm. 
During this step, the connected cluster containing $i$ is determined. Let us call its size 
$C_n$, whence
\be 
      S_n = \max\{S_{n-1},C_n\}.
\ee
The $n$th gap is defined as 
\be
     \Delta_n = S_{n+1} - S_n,
\ee
and the maximal gap over all values of $n$ is called $\Delta_{\rm max}$, while the 
$n$-value at which the maximum occurs is called $n_{\rm max}$ and the giant cluster size 
at this point is $S_{\rm max}$.

As observables we measured the average order parameter and its variance as functions of $n$,
the averages of $\Delta_{\rm max}$ and $n_{\rm max}$, and their variances. 
These were measured at lattice sizes $L=32, 64, \ldots 16384$ for $d=2$, and at $L=32,64,\ldots
512$ for $d=3$. The number of realizations for each Levy flight parameter $\alpha$ and for 
each dimension in the case of (generalized) Knight's move RWs was $> 70\,000$ for the largest
$L$, and increased up to $>2\,000\,000$ for the smallest.

\section{Finite-size scaling}


Because FSS might be different in the present model in view of the additional length scale 
induced by the finiteness of the walk time $T$, we should review the standard scenario for
its scaling.

We expect that ${\rm Var}[S_n]$ has a peak near the percolation transition which gets sharper 
with increasing $N$. At the same values of $N$, the gaps should also be maximal. Let us call 
$\rho_c(L)$ the position of the peak of the distribution of $n_{\rm max}/N$ at given $N=L^d$, 
and $\rho_c = \lim_{L\to \infty}$. Let us furthermore define the order parameter exponent 
$\beta$ and the correlation length exponent $\nu$ by demanding for infinite systems that
\be
   s \equiv L^{-2} \langle S_n(\rho)\rangle \sim (\rho-\rho_c)^\beta \quad {\rm for} \rho > \rho_c 
   \label{beta;S}
\ee 
and 
\be
    \xi(\rho) \sim |\rho-\rho_c|^{-\nu}\;,                                      \label{nu;chi}
\ee
where $\xi(\rho)$ is the correlation length which for percolation is defined as the rms
radius of the largest finite cluster.

Standard (FSS) arguments (mainly that observables are homogeneous functions near a critical 
point, that there is only one unique divergent length scale as $\rho \to \rho_c$, and that the 
scaling of a quantity depends only on its (anomalous) dimension, lead to the ansatzes
\be
   s = L^{d_f-d}\Psi_S[(\rho-\rho_c) L^{1/\nu}]
   \label{FSS-s}
\ee
and
\be
   \chi \equiv L^{-2}\{{\rm Var}[S_n(\rho)]\}^{1/2}  = L^{d_f-d}\Psi_\chi[(\rho-\rho_c) L^{1/\nu}],
   \label{FSS-chi}
\ee
where 
\be
   d_f = d-\beta/\nu.     \label{df-beta}
\ee
For {\it bond} percolation (whether correlated or not), $S_n$ would increase whenever the largest 
cluster eats a smaller one. The largest gap would thus occur when the largest second-largest 
cluster gets eaten. If we still assume that all masses scale with $L$ according to their anomalous 
dimension, this would imply that also 
\be
   \langle \Delta_{\rm max}\rangle \sim \chi_{_\Delta} \;
                      \equiv\; \{{\rm Var}[\Delta_{\rm max}]\}^{1/2} \sim L^{d_f}
   \label{Delta_scaling}
\ee
at criticality, while equations analogous to Eqs.~(\ref{FSS-s}) and (\ref{FSS-chi}) (with scaling 
functions $\Psi_\Delta$ and $\Psi_{\chi_{_\Delta}}$) should hold for $\rho \neq \rho_c$.

For the present case of site percolation, essentially the same argument applies. There, $\Delta_n$ 
corresponds to the sum of a small number of eaten neighboring clusters, and Eq.(\ref{Delta_scaling})
can be assumed still to hold. 

Finally, we expect that distributions of observables like $S_{\rm max}, \rho_{\rm max}$ 
(the density of visited sites where the largest gap occurs) and $\Delta_{\rm max}$ should 
be, up to normalization, functions of dimensionless variables, where we can write 
$S_{\rm max}/L^{d_f}, (\rho-\rho_c)L^{1/\nu}$ and $\Delta/L^{d_f}$, so that we can write
\be
   P_S(S_{\rm max}) = L^{-d_f} f_S(S_{\rm max}/L^{d_f}),       \label{P_S}
\ee
\be
   P_{\rho}(\rho_{\rm max}) = L^{1/\nu} f_\rho[(\rho_{\rm max}-\rho_c)L^{1/\nu}]  \label{P_rho}
\ee
and 
\be
   P_{_\Delta}(\Delta_{\rm max}) = L^{-d_f} f_{_\Delta}(\Delta_{\rm max}/L^{d_f})  \label{P_Delta}
\ee
(notice that Eqs. (11) and (13) of Ref. \cite{Fan}, which are analogous to Eqs.(\ref{P_S}) and
(\ref{P_Delta}) are more complicated without need).

According to the standard FSS scenario, the variance of $S_n$ and the distribution of 
$\rho_{\rm max}$ have near-by peaks which have the same scaling with $L$ and whose 
position is shifted from $\rho_c$ by the same scaling. If we denote the average of these
two peak positions as $\rho_c(L)$, we should thus have
\be
   \rho_c(L) -\rho_c \sim {\rm peak}\;{\rm width} \sim L^{-1/\nu}.
\ee

\section{Numerical Results}

\subsection{Two dimensions}

\subsubsection{Conventional variables}

\begin{figure}
  \begin{center}
  \psfig{file=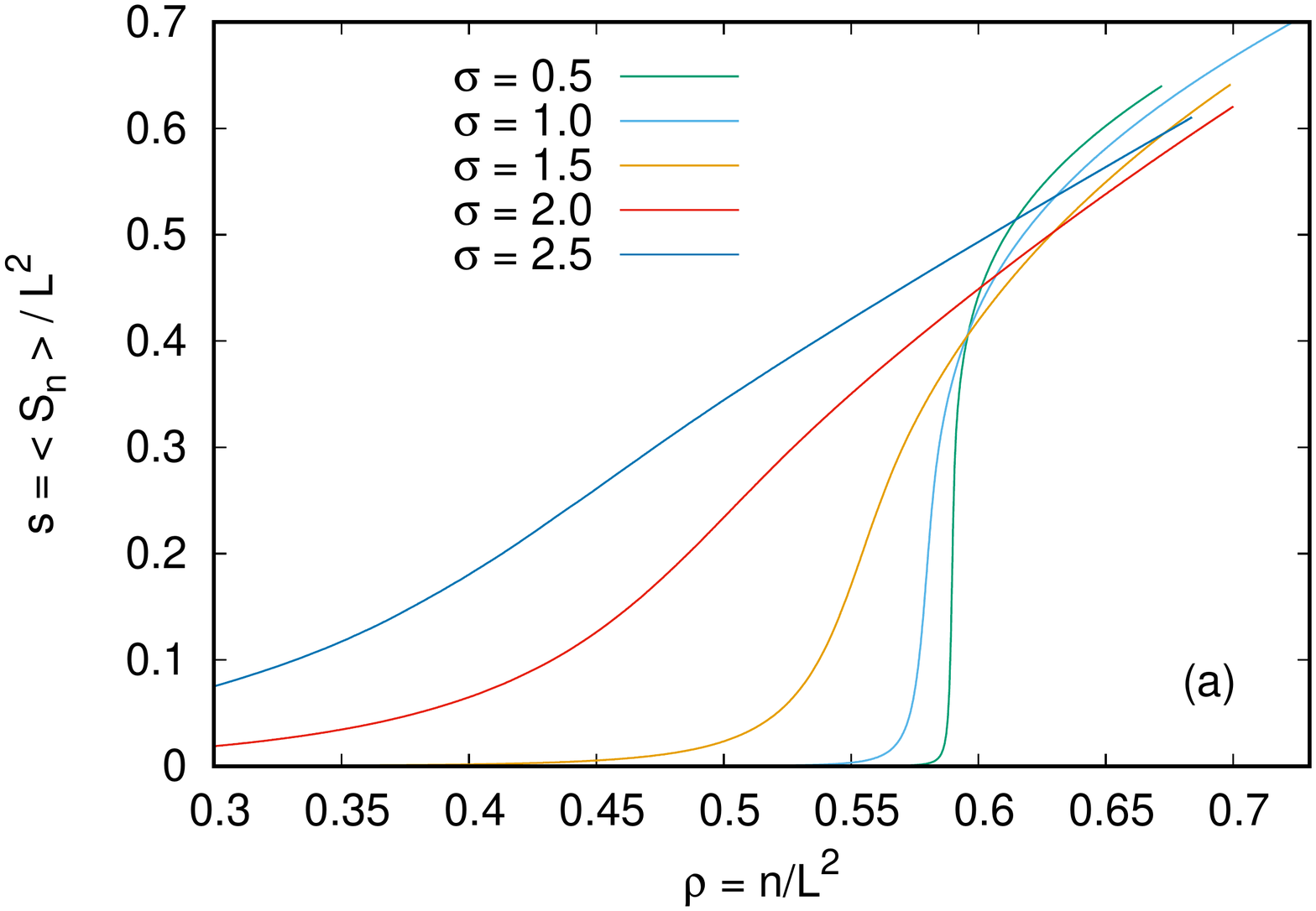,width=8.5cm, angle=0}
  \vglue -17pt
  \psfig{file=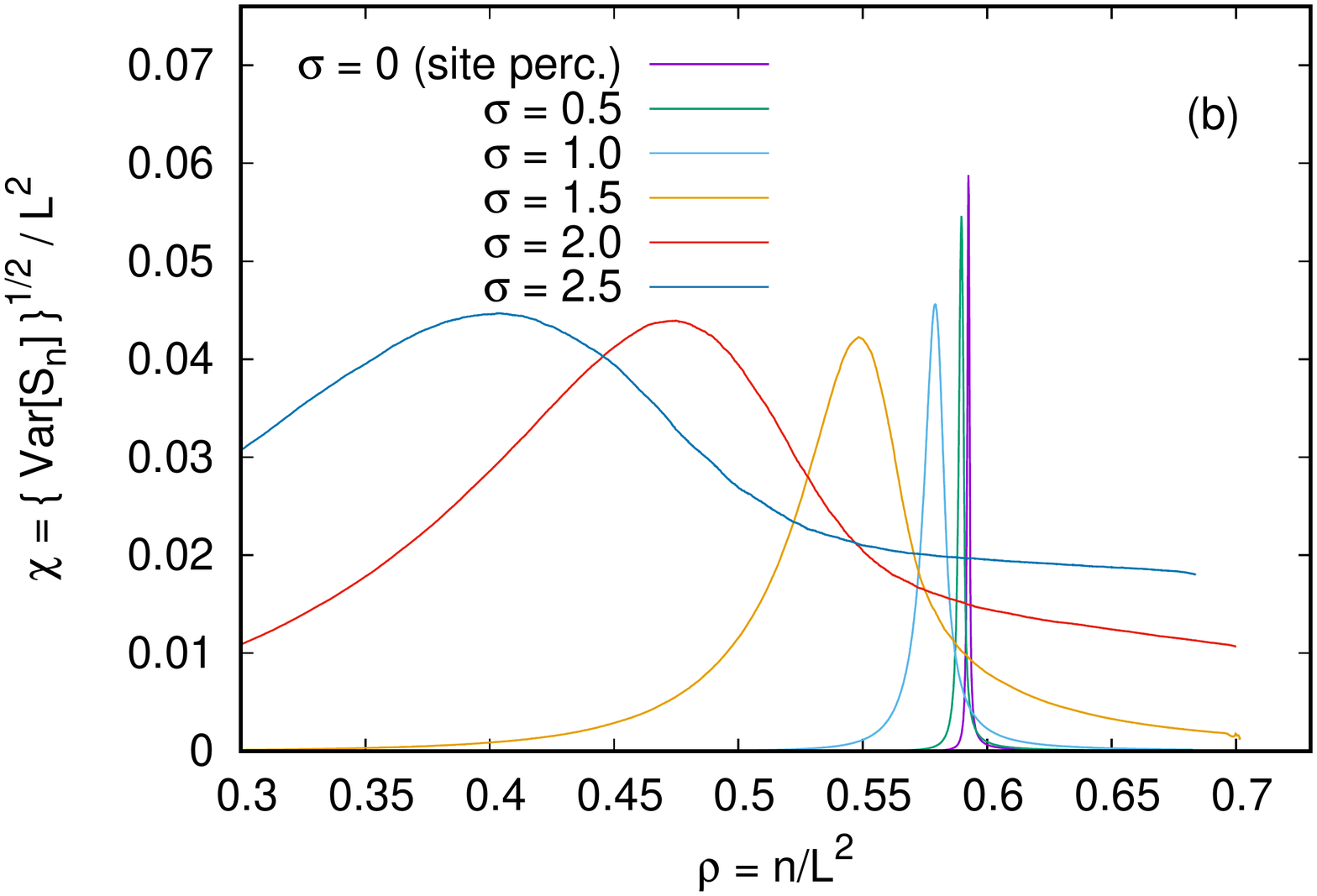,width=8.5cm, angle=0}
  \vglue -17pt
  \caption{Order parameters (a) and the square root of their variances (b) at $L=16384$,
     for five values of $\sigma$ between 0 and 2.5, plotted against the density of allowed
     sites $\rho$ which serves as a control parameter.}
    \label{overview}
  \end{center}
  \vglue -7pt
\end{figure}

We studied percolation on the sites visited by Levy flights with $\sigma = 0.1,0.2,0.3,0.5,0.75,
1.0, 1.25,1.5,1.7$, $1.8,1.9,2.0$ and $2.5$. The  last two values are strictly speaking no longer Levy
flights (where $\sigma < 2$ for $d=2$) but scale like ordinary RWs, but we also can  use the Levy flight
generating algorithm for these values, and get nontrivial results because the visited sites
do not form, in general, connected clusters. We also simulated ordinary site percolation, which
corresponds to $\sigma =0$, to see whether the scaling changes when going from $\sigma =0$
to $\sigma >0$.

In Fig.~\ref{overview} we show the order parameter $s$ and its fluctuations $\chi$ as functions of 
the density $\rho$ of visited sites, for $N=16384$ and for typical values of $\sigma$. We see
the very sharp transition for ordinary site percolation ($\sigma = 0$), while the transitions 
become increasingly more fuzzy for increasing $\sigma$ and happen at smaller densities of allowed sites. 
Indeed we claim that the leftmost curve (for $\sigma=2.5$) 
and maybe also that for $\sigma = 2$ do not show phase transitions at all. To settle this 
question, we also have to look  at smaller $L$ and perform careful FSS analyses.

\begin{figure}
  \begin{center}
  \psfig{file=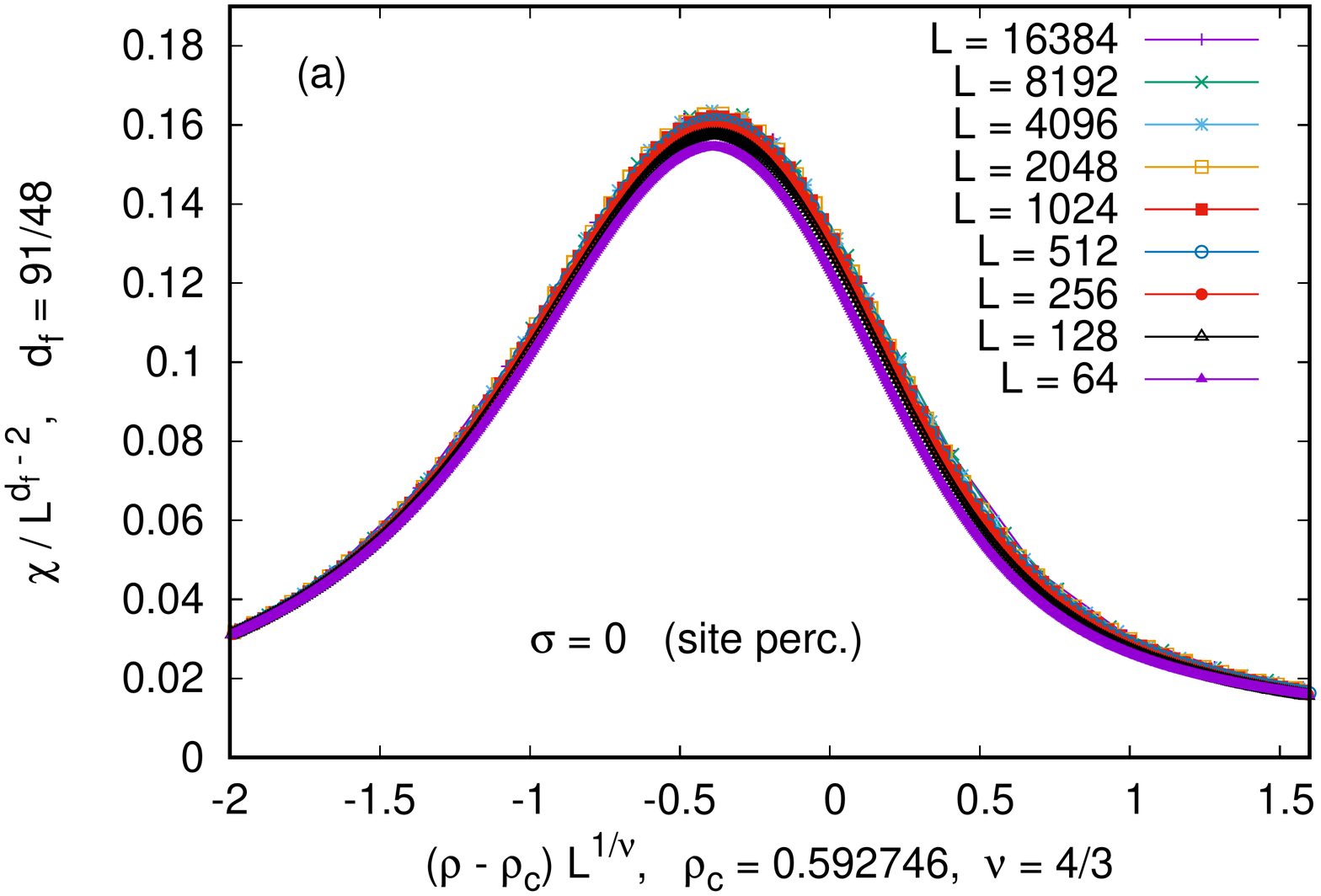,width=8.5cm, angle=0}
  \vglue -17pt
  \psfig{file=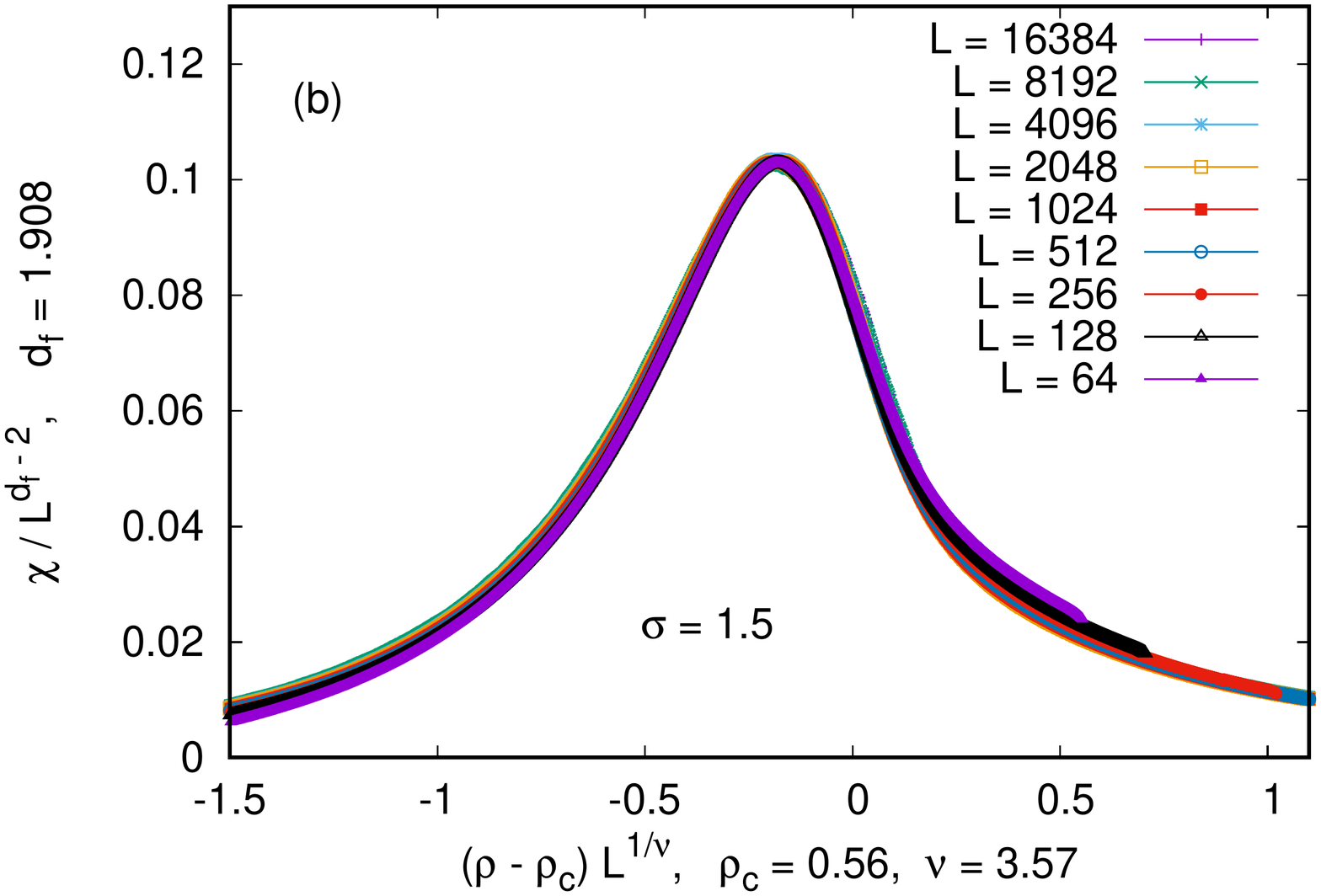,width=8.5cm, angle=0}
  \vglue -17pt
  \caption{Data collapse plots of $\chi$ against $\rho$ for $\sigma = 0$ (panel a) and 
     $\sigma = 1.5$ (panel b). The critical exponents used in these plots are the exact ones
     for standard OP in panel (a), and fitted ones in panel(b). Note that, in view of the 
     visible deviations from a perfect data collapse in panel (a) (where the asymptotic scaling 
     is known exactly), the good data collapse in panel (b) might be a bit fortuitous, and 
     the precise values of the fitted exponents for $\sigma = 1.5$ should not be taken too 
     seriously.}
    \label{chi}
  \end{center}
  \vglue -7pt
\end{figure}

In Fig.~\ref{chi} we show the values of $\chi$ against $\rho$ for $L$ ranging from 64 to 16384,
and for $\sigma = 0$ (panel a) and $\sigma = 1.5$ (panel b). More precisely, in view of 
Eq.~(\ref{FSS-chi}), we plotted $L^{d-d_f} \chi$ against $(\rho-\rho_c)L^{1/\nu}$, where we 
took the standard OP values of $d_f$ and $\nu$ for $\sigma=0$, but had to use fitted values
of the critical exponents for $\sigma = 1.5$.  There are several comments: 

(i) The collapse is not perfect even for $\sigma = 0$ (where we know the exact asymptotic 
scaling), which illustrates the importance of non-leading corrections to scaling. This  also
shows that using least-squares fits to obtain the best data collapse in such 
figures could be highly misleading. Indeed, data collapse plots like Fig.~\ref{chi} are 
very helpful in getting rough overviews, but other methods are, in general, better suited 
to obtain precise results. For percolation, these include, e.g., spanning probabilities 
\cite{Ziff}, the mass of the second-largest cluster at criticality \cite{Margolina}, 
or the scaling of gaps as discussed in the previous section \cite{Manna,Nagler,Fan}. 
In the present case, estimating spanning probabilities or second-largest
cluster masses would abrogate the advantages of the NZ algorithm, and was thus not done. 

(ii) With increasing $\sigma$, the fractal dimension $d_f$ increases slightly, but it hardly 
changes. In contrast, $\nu$ increases dramatically. But we still obtain a perfect data collapse,
which implies that the width of the peak and its shift from the exact critical point (which has
also decreased significantly from its value for $\sigma = 0$) scale in the same way with $L$.
Thus we see here no indication for two different $\nu$-exponents.

\begin{figure}
  \begin{center}
  \psfig{file=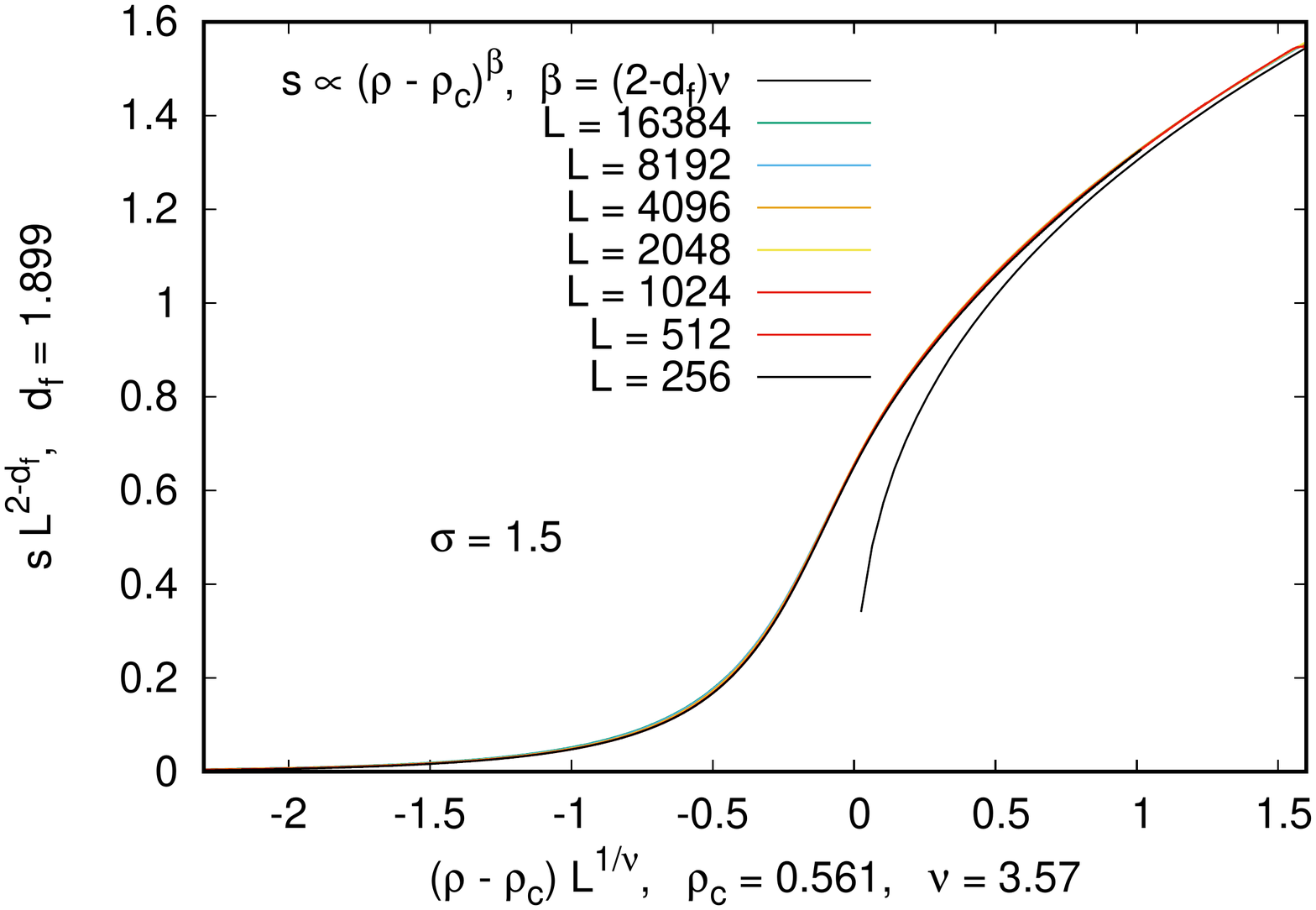,width=8.5cm, angle=0}
  \vglue -17pt
  \caption{Data collapse plot of $s$ against $\rho$ for $\sigma = 1.5$. The exponent $\nu$
     is the same as in Fig.~\ref{chi}b, but $d_f$ and $\rho_c$ are slightly re-adjusted for 
     best collapse. Also plotted is a power law $s = const \, (\rho-\rho_c)^\beta$, 
     showing that Eqs.~(\ref{beta;S}) and (\ref{df-beta}) are well satisfied.}
    \label{beta-1.5}
  \end{center}
  \vglue -7pt
\end{figure}

The critical threshold $\rho_c$ and the exponents $\nu$ and $\beta$ can also be estimated by 
using Eqs.~(\ref{beta;S}) and (\ref{FSS-s}). In Fig.~\ref{beta-1.5} we show, again for $\sigma=1.5$,
a data collapse plot in which we plotted $L^{2-d_f}s$ against $(\rho-\rho_c)L^{1/\nu}$. We used 
the same value of $\nu$ as in Fig.~\ref{chi}b, but for optimal collapse we had to use slightly 
different values of $d_f$ and $\rho_c$. Since precise error estimates are difficult from such 
data collapse plots, we see these differences as rough error estimates. In addition, we show in 
Fig.~\ref{beta-1.5} a curve indicating $const \, (\rho-\rho_c)^\beta$, with $\beta = (2-d_f)\nu$.
It shows that Eqs.~(\ref{beta;S},\ref{df-beta}) are rather well satisfied.

Similar analyses were also made for other values of $\sigma$, but we do not report results
since more precise estimates of critical parameters are obtained from gap scaling, as we
shall show next.

\subsubsection{Gap scaling in the event-based ensemble}

In the above conventional types of analyses, observables are studied at fixed values of the 
control parameters. It was suggested first by Manna and Chatterjee \cite{Manna} (see also 
Refs. \cite{Nagler,Fan,Feshanjerdi,Li2023}) that more precise estimates could be obtained by studying 
observables at that value of the control parameter where the largest gap (i.e., the largest 
jump in the order parameter) occurs in individual realizations. These values fluctuate of course 
from realization to realization, and the ensemble of realizations at the point of maximal gap
is called event-based ensemble in Ref. \cite{Li2023}. This was proposed for EP  \cite{Manna,Nagler}, where these fluctuations are excessively large \cite{Christensen}, 
and its usefulness for other percolation transitions was suggested in Refs. \cite{Fan,Feshanjerdi}.

\begin{figure}[h]
  \begin{center}
  \psfig{file=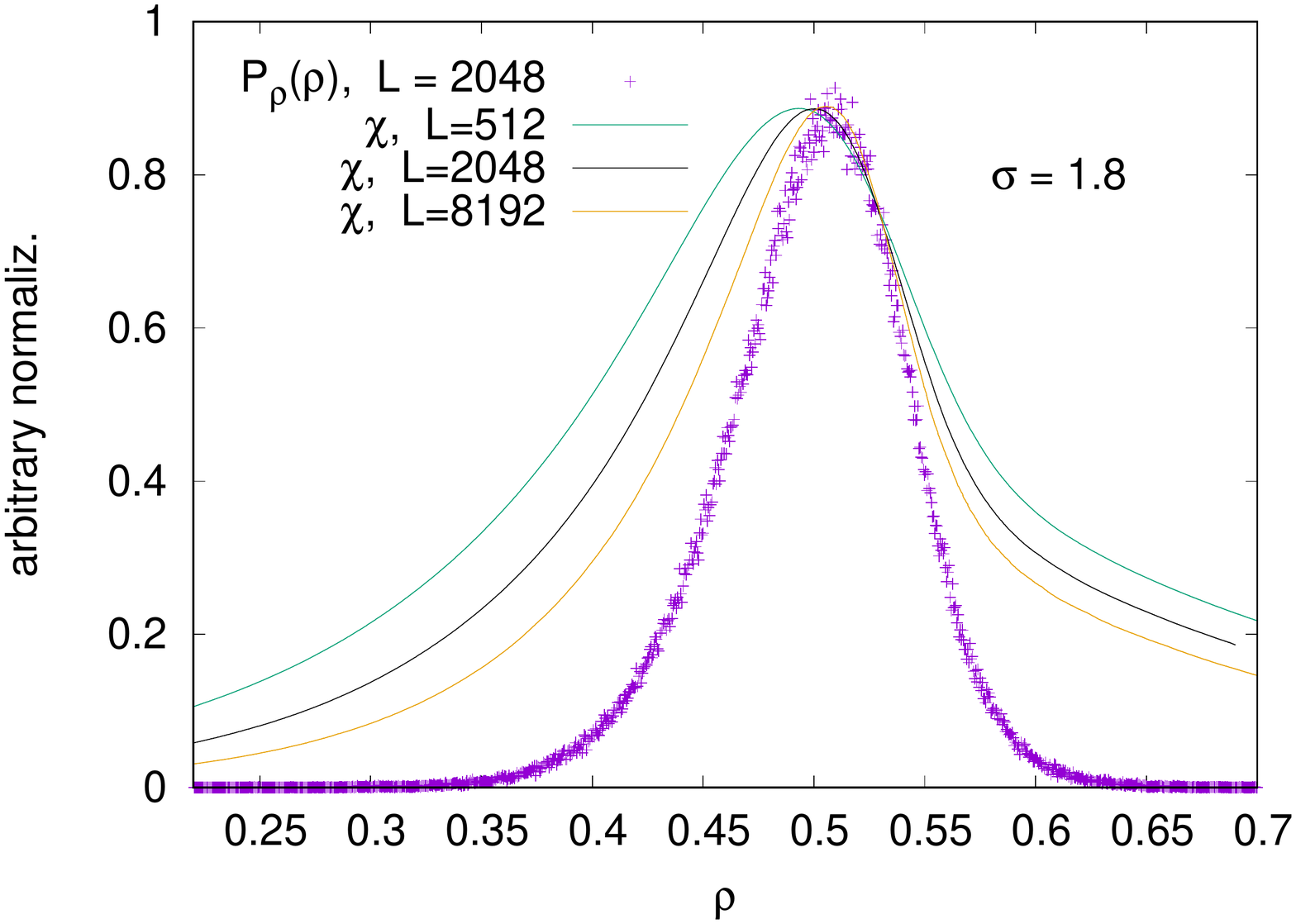,width=8.5cm, angle=0}
  \vglue -17pt
     \caption{Plots of $P_\rho(\rho)$ (the distribution of maximal gap positions) and of the 
     width $\chi$ of the order parameter distribution at given $\rho$ at $\sigma=1.8$. 
     Normalization of all curves is such that they all have the same height, for easier comparison 
     of their widths. It is seen that $P_\rho(\rho)$ has the sharpest peak, even if we compare it 
     to curves of $\chi$ at different values of $L$.}
    \label{peaks-18}
  \end{center}
  \vglue -7pt
\end{figure}

That gap scaling studied at the points of maximal gaps is also useful in the present model is 
suggested by Fig.~\ref{peaks-18}. There we plotted $P_\rho(\rho)$ (the distribution of maximal 
gap positions) at $L=2048$ and $\sigma=1.8$, and compared it to three curves of $\chi$ at the same 
value of $\sigma$ and for three different values of $L$. For easier comparison of their widths, we 
used the same arbitrary normalization for all four curves. It is clearly seen that $P_\rho(\rho)$ 
has the narrowest peak. It has the largest fluctuations, but this drawback is far outweighed
by the sharpness of its peak.

\begin{figure}[h]
  \begin{center}
  \psfig{file=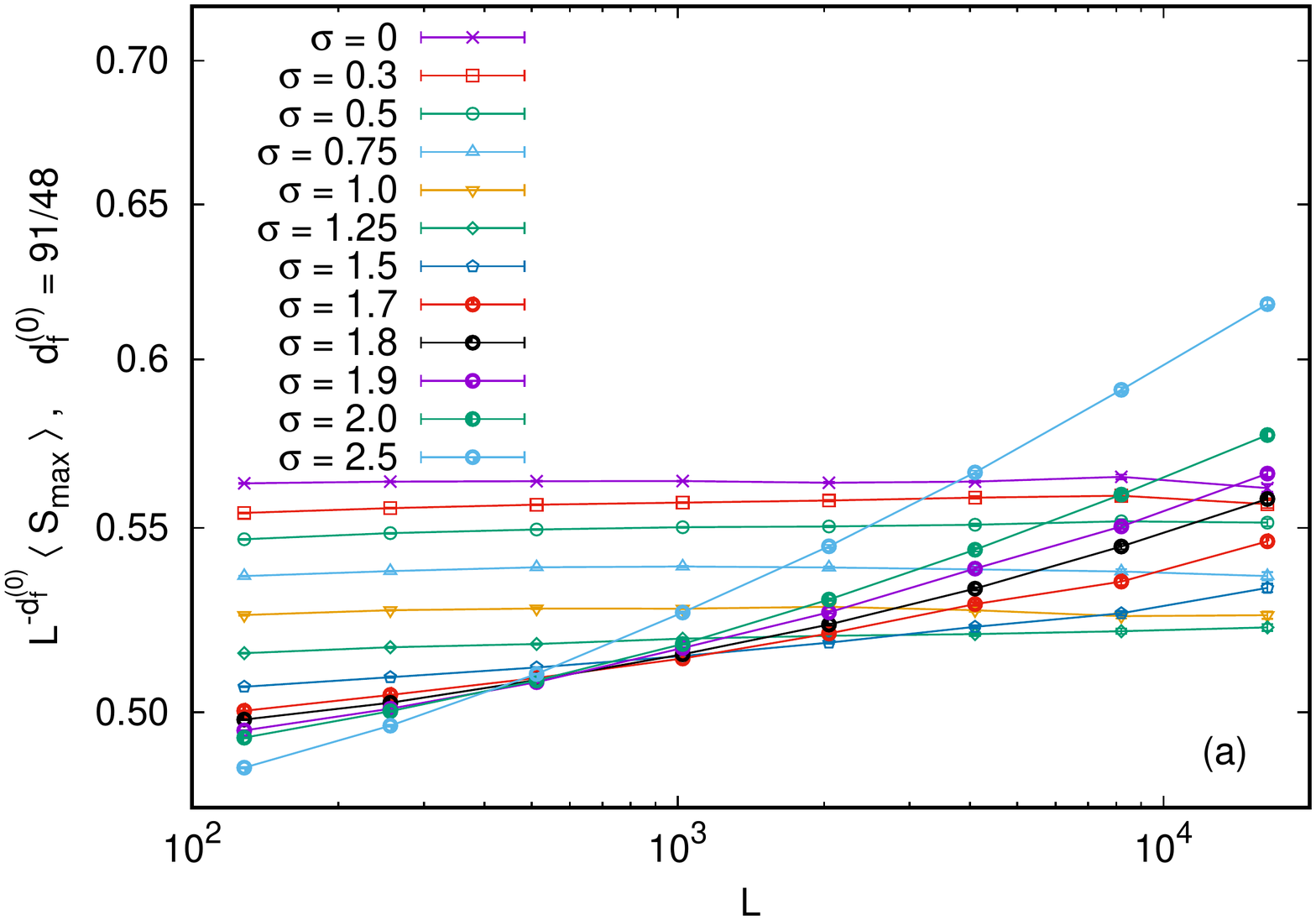,width=8.5cm, angle=0}
  \vglue -17pt
  \psfig{file=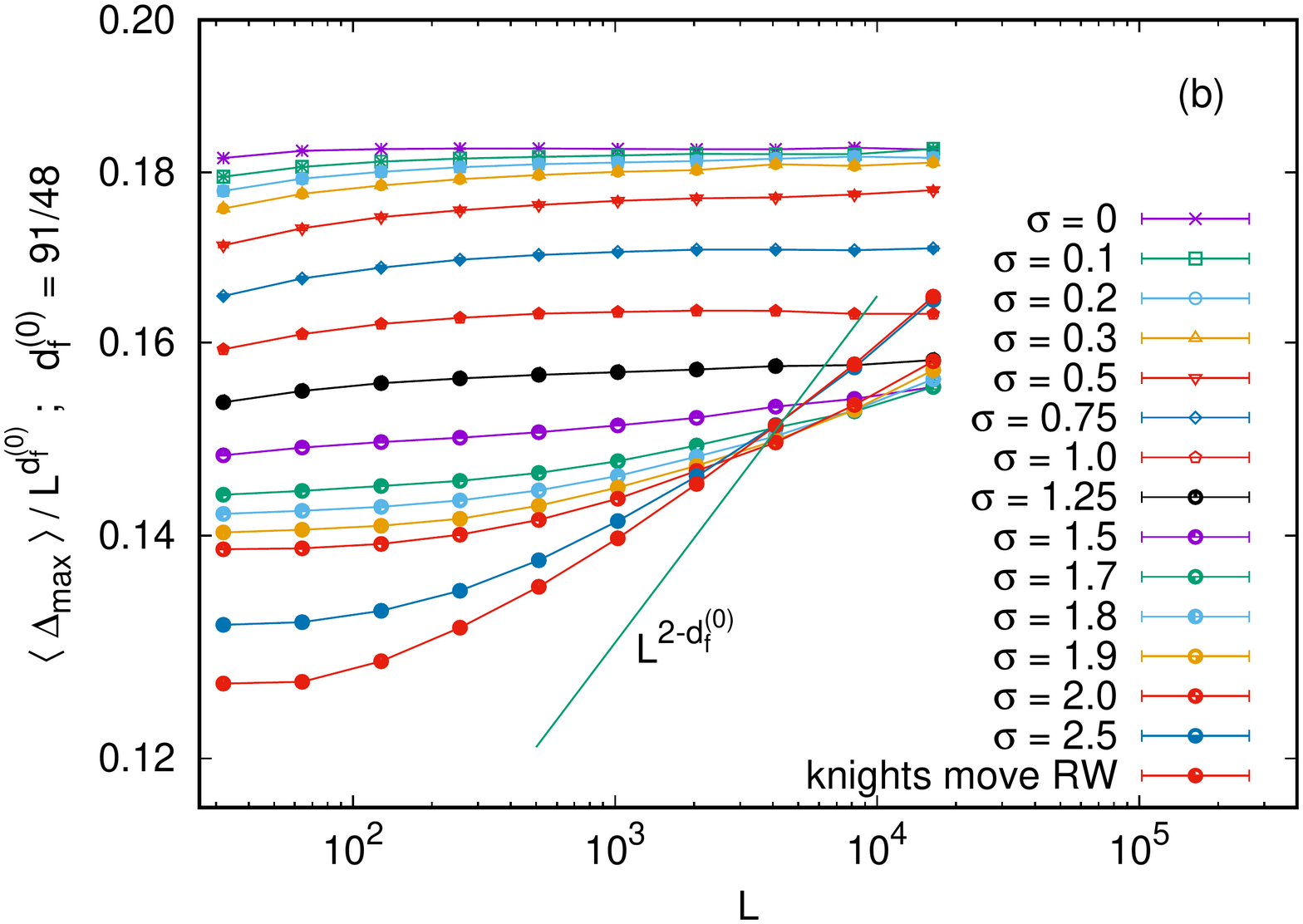,width=8.5cm, angle=0}
  \vglue -17pt
     \caption{Log-log plots of $L^{-d_f^{(0)}} \langle S_{\rm max}\rangle$ (panel a) and of
     $L^{-d_f^{(0)}} \langle \Delta_{\rm max}\rangle$ (panel b) against $L$. In panel b we also 
     show a straight line with the slope that would be expected for compact clusters ($d_f = 2$).}
    \label{df-averages}
  \end{center}
  \vglue -7pt
\end{figure}

\begin{figure}[h]
  \begin{center}
  \psfig{file=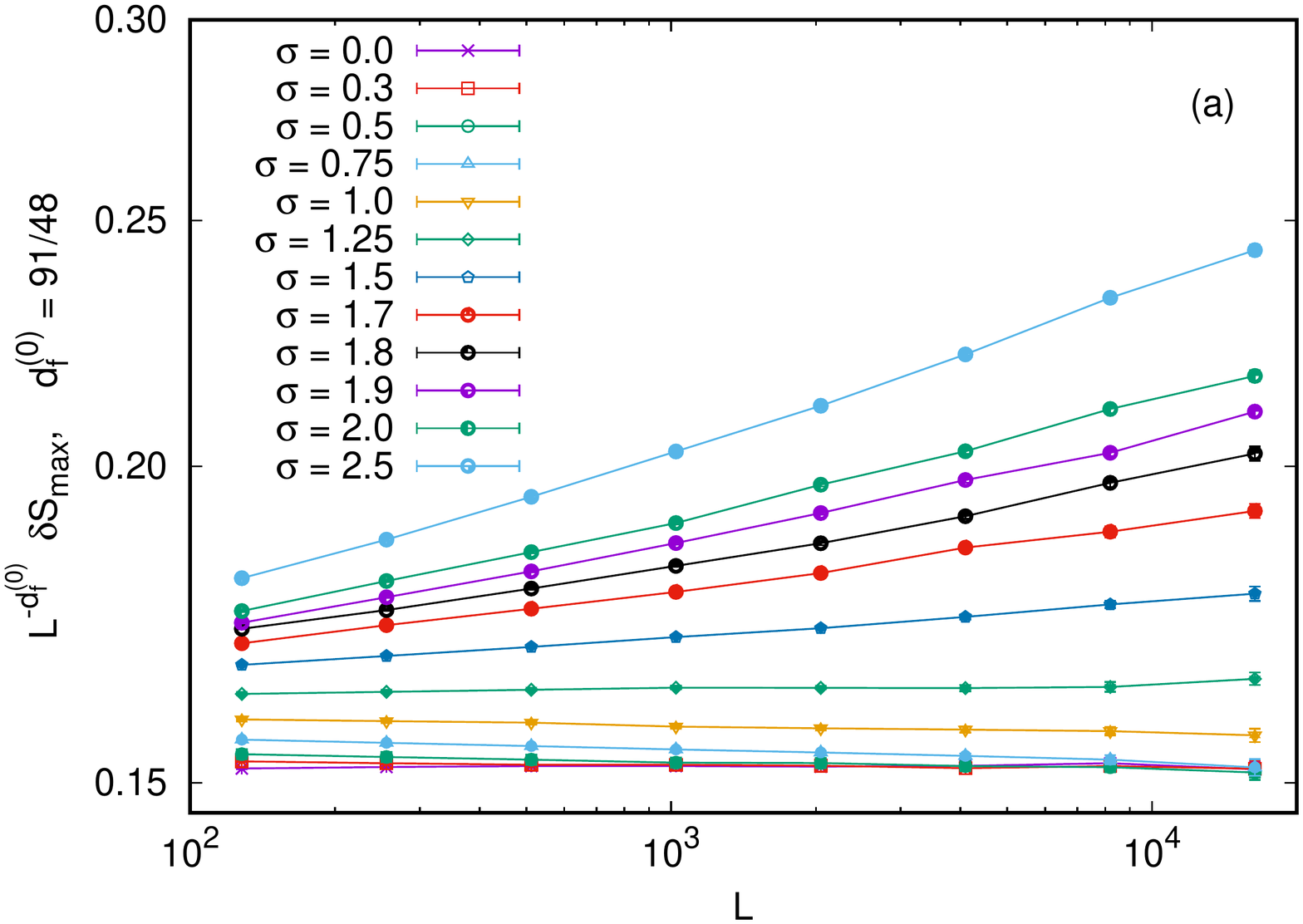,width=8.5cm, angle=0}
  \vglue -17pt
  \psfig{file=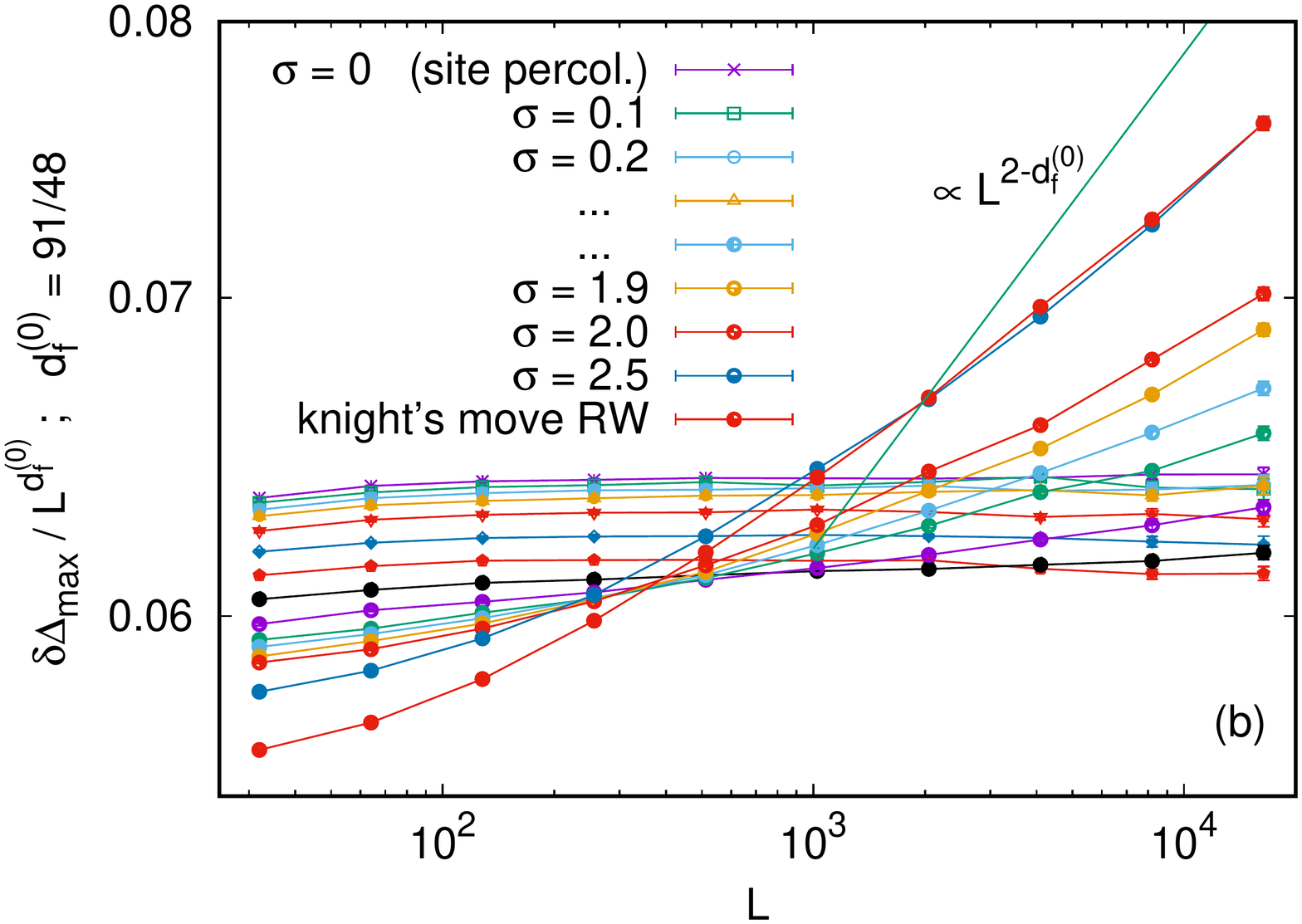,width=8.5cm, angle=0}
  \vglue -17pt
     \caption{Log-log plots analogous to Fig.~\ref{df-averages}, but of $\delta S_{\rm max} = 
     \{{\rm Var}[S_{\rm max}]\}^{1/2}$ (panel a) and $\delta \Delta_{\rm max} =
     \{{\rm Var}[\Delta_{\rm max}]\}^{1/2}$ (panel b), instead of $\langle S_{\rm max}\rangle$
	  and $\langle \Delta_{\rm max}\rangle$.}
    \label{df-variances}
  \end{center}
  \vglue -7pt
\end{figure}

{\bf Fractal dimensions:}
Let us first look at the fractal dimensions. It can either be obtained from the average values
and variances of $S_{\rm max}$ (the size of the giant cluster at criticality), or from the 
average values and variances of $\Delta_{\rm max}$ (which, as we pointed out, should scale like
the size of the second-largest cluster). In Fig.~\ref{df-averages} we show log-log plots of
$L^{-d_f^{(0)}} \langle S_{\rm max}\rangle$ (panel a) and of
$L^{-d_f^{(0)}} \langle \Delta_{\rm max}\rangle$ (panel b) against $L$, where $d_f^{(0)} = 91/48$
is the fractal dimension in OP. We see in both panels that the curves are horizontal for 
$\sigma < 1$, suggesting that the model is in the OP universality class for $\sigma < 1$.
For $\sigma > 1$ there are, however, significant deviations which become more and more pronounced
with increasing $\sigma$. But since all curves are strongly non-linear, it is impossible to quote
with certainty an asymptotic power law for any $\sigma > 1$. We also indicate in both panels 
the power laws $S_{\rm max} \sim \Delta_{\rm max} \sim L^2$, which we would expect for compact
clusters. It is very strongly suggested that this is the asymptotic scaling for $\sigma>2$ (and 
for Knight's move RWs), 
and we will later give strong arguments that there is no sharp percolation transition in this 
case. Whether there is a sharp transition for $\sigma=2$ is an open question.

Analogous plots for the (square roots of the) variances are shown in Fig.~\ref{df-variances}.
Again both panels of Fig.~\ref{df-variances} clearly show OP scaling for $\sigma <1$, and 
non-OP scaling for $\sigma > 1$. But again it is impossible to determine the asymptotic
scaling laws for $\sigma > 1$, except that the data suggest strongly that clusters are compact
for Levy flights with $\sigma>2$ and for the Knight's move RWs.

All these results agree perfectly with what we obtained from the conventional analysis (data
not shown). In particular, we understand now that the fractal dimensions used in 
Figs.~\ref{chi}b and \ref{beta-1.5} are only effective exponents valid in the studied
range of $L$, and it should not surprise that they differ from each other.

\begin{figure}[t]
        \centering
	\psfig{file=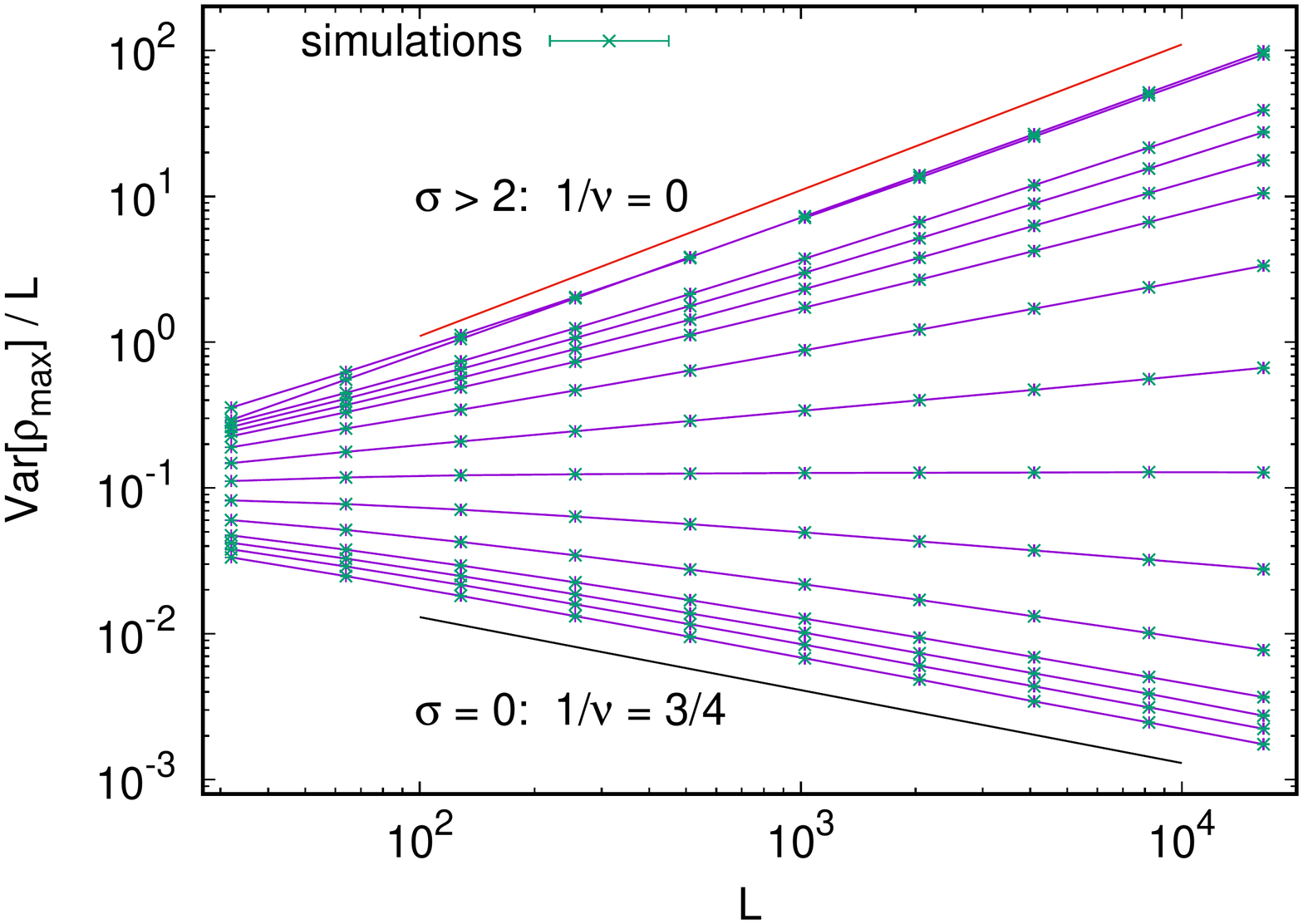,width=8.5cm, angle=0}
        \caption{Plots of the variances of times of largest jumps as a function of $L$ for 
	$\sigma = 0.0, 0.1, 0.2, 0.3, 0.5, 0.75, 1.0, 1.25, 1.5, 1.7, 1.8, 1.9, 2, 2.5$, 
	and Knight's move RW from the bottom to the top of the curves, respectively. 
	The upper solid line corresponds to $\nu=\infty$ and seems to apply for $\sigma > 2$ and Knight's move RW,
	while the lower line corresponds to $\nu=4/3$ which holds for standard $2D$ percolation, and is consistent 
	with our results within error bars for all $\sigma < 0.5$.}
        \label{chi-rho}
\end{figure}

{\bf Correlation length exponents:}
Correlation length exponents are obtained from the scalings of the shift of the averages of 
$\rho_{\rm max}$ and of the widths of their distributions. According to standard FSS, both
give the same exponent $\nu$, but due to possible violations of the standard FSS scenario, 
this might not be the case in the present model.

Since measuring the shifts of $\bar{\chi} \equiv \langle \rho_{\rm max}\rangle$ with $L$ requires precise 
estimates of the true critical point positions, this is a somewhat delicate and error-prone
procedure, in particular since we have already seen strong deviations from pure power law
scalings. Thus we look first at the scaling of the variances. In Fig.~\ref{chi-rho} we show
log-log plots of $L^{1/2} \chi_\rho$ against $L$, where 
\be
    \chi_\rho = \{{\rm Var}[\rho_{\rm max}]\}^{1/2}\;.
\ee

We see now strong deviations from OP scaling for all $\sigma > 0.5$. Superficially, all curves 
look rather straight so that $\nu$ seems well determined for each $\sigma > 0.5$ and $1/\nu$ 
seems to increase continually with it, until $1/\nu =0$ for $\sigma > 2$
(which would suggest that $\chi_\rho = const$ for $\sigma > 2$). But more careful inspection
shows that all curves for $\sigma<1$ bend downwards, while those for $\sigma>1$ bend up. Only
the curve for $\sigma=1$ seems perfectly straight for $L>256$, with slope 
\be
    \nu^{(\sigma=1)} = 2.00\pm 0.03\;.
\ee
It is not clear what this means for the true asymptotic values of $\nu$. If the deviations from
straight lines are a minor finite size correction (which is suggested superficially), then
$1/\nu$ seems to decrease roughly linearly with $\sigma$ in the range $1/2 < \sigma  < 2$, i.e.,
\be
   1/\nu = \left\{\begin{array}{r@{\quad:\quad}l}
	              3/4        & \sigma < 1/2 \\
 	              1-\sigma/2 & 1/2 < \sigma  < 2 \\
	              0          & \sigma > 2
                  \end{array} \right.
\ee
This would mean that the model is not in the OP class for $1/2 < \sigma  <1$, although we had 
clear evidence that $d_f$ there is the same as in OP.

Another, more radical, extrapolation could be the following: The curvatures seen in Fig.~\ref{chi-rho} 
imply that all curves for $\sigma <1$ align asymptotically with the one for $\sigma =0$, and those
for $\sigma >1$ become finally parallel to that for $\sigma=2$. In this scenario, $\nu$ is would 
be constant for all $\sigma\neq 1$, and that it jumps at $\sigma = 1$ from 
$4/3$ to $\infty$. Neither of these two scenarios is very plausible. A third one could be 
that $1/\nu=1/\nu^{(0)}$ for $\sigma < 1$, and decreases then continuously to 0.

Whatever the correct scenario is, it is clear that $1/\nu = 0$ for $\sigma >2$, which means
that the order parameter curve $s$ versus $\rho$ becomes, for $\sigma >2$, independent of $L$, 
and in particular no singularity develops in the limit $L\to\infty$. Thus there is no 
percolation transition for $\sigma >2$.

\begin{figure}[h]
  \begin{center}
  \psfig{file=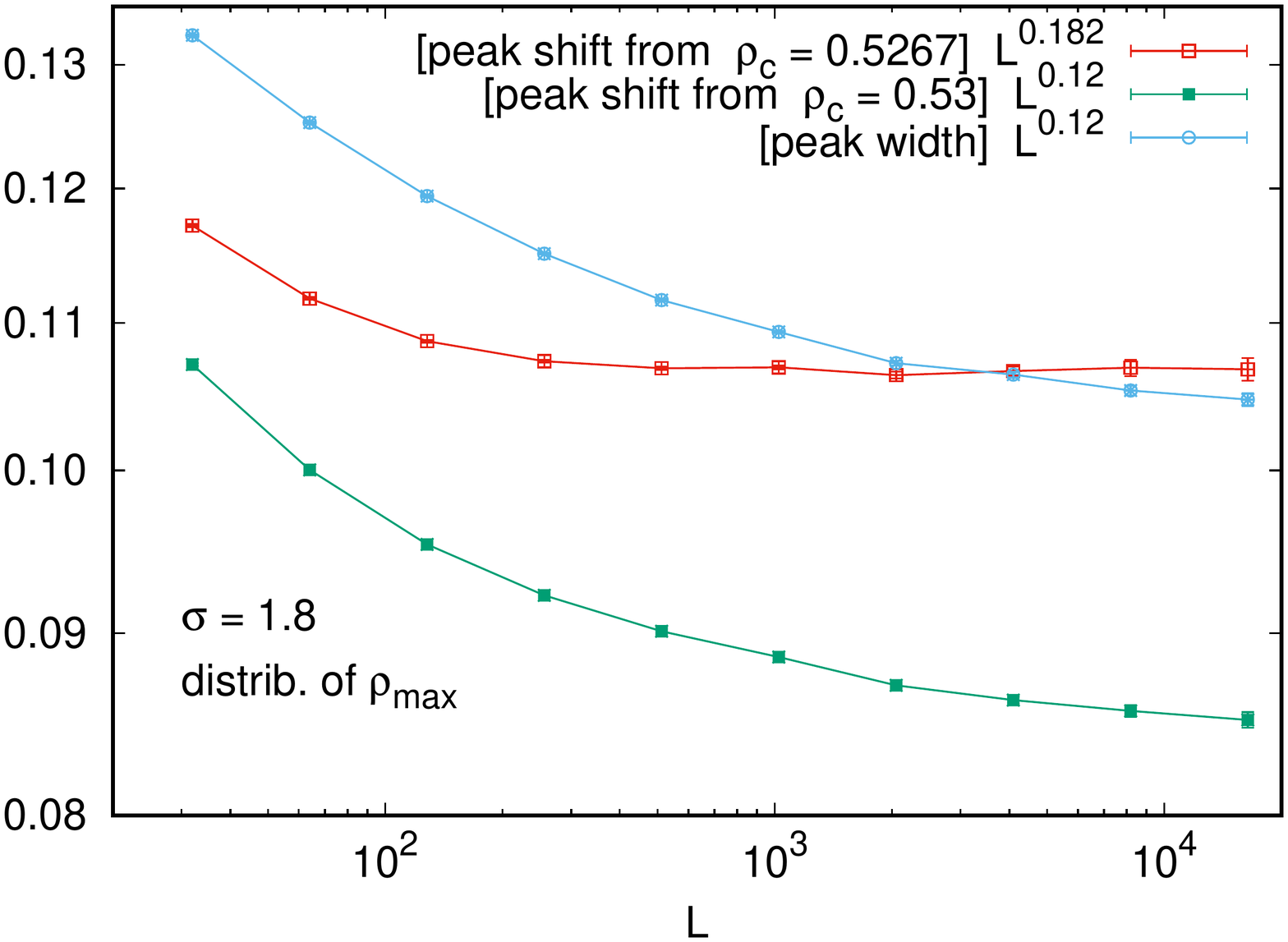,width=8.5cm, angle=0}
  \vglue -17pt
  \caption{Log-log plots of $L^{1/\nu} \chi_\rho$ and of $(\bar{\rho} - \rho_c) \, L^{1/\nu}$ 
     versus $L$ for $\sigma=1.8$. For $\chi_\rho$ we choose $\nu$ such that the curve seems
     to become flat for large $L$. For $\bar{\rho} - \rho_c$ we show two curves: One, where
     the curve shows best scaling (for all $L>256$), the other with the same $\nu$ as for 
     $\chi_\rho$ and with $\rho_c$ such that it becomes nearly a shifted copy of the 
     one for $\chi_\rho$.}
    \label{nu-1.8}
  \end{center}
  \vglue -7pt
\end{figure}

Let us now look at the values of $\bar{\rho}$ and their dependences on 
$\sigma$ and $L$. To be specific, take $\sigma = 1.8$. In Fig.~\ref{chi-rho} we had seen that
if there is a scaling law $\chi_\rho \sim L^{1/\nu}$, then there must be very large 
finite size corrections to it. In contrast, if we choose $\rho_c(\sigma = 1.8)$ carefully,
we can make the curve of $\log (\bar{\rho} - \rho_c(\sigma = 1.8))$ versus $\log L$ nearly 
perfectly straight -- but with a value of $\nu$ which is closer to $L^{1/\nu^{(0)}}$.
This would support the conjecture that there are two different correlation length exponents.
But there is also another, more plausible scenario: If we allow similarly large corrections
to scaling for the dependence of $\bar{\rho}$ on $L$ as for $\chi_\rho$, we can find a value
of $\rho_c$ such that the curves $\bar{\rho} - \rho_c$ versus $L$ and  $\chi_\rho$ versus
$L$ give practically the same value of $\nu$. This is demonstrated in Fig.~\ref{nu-1.8},
where we plotted both quantities against $L$ with suitably chosen values of $\nu$ and 
$\rho_c$. More precisely, in this log-log plot we show one curve for $\chi_\rho$ and 
two curves for $\bar{\rho} - \rho_c$ -- one such that is it as straight as possible, the 
other such that it mimics $\chi_\rho$.

We thus conclude  that the model definitely is not in the OP universality class for $\sigma > 1$.
The possible deviations from the conventional FSS picture due to a possible new length scale
generated by the finite times of the Levy flights seem  not to have led to two values of $\nu$,
but they might be the source for the huge observed corrections to scaling.

\subsection{Three dimensions}

\begin{figure}
  \begin{center}
  \psfig{file=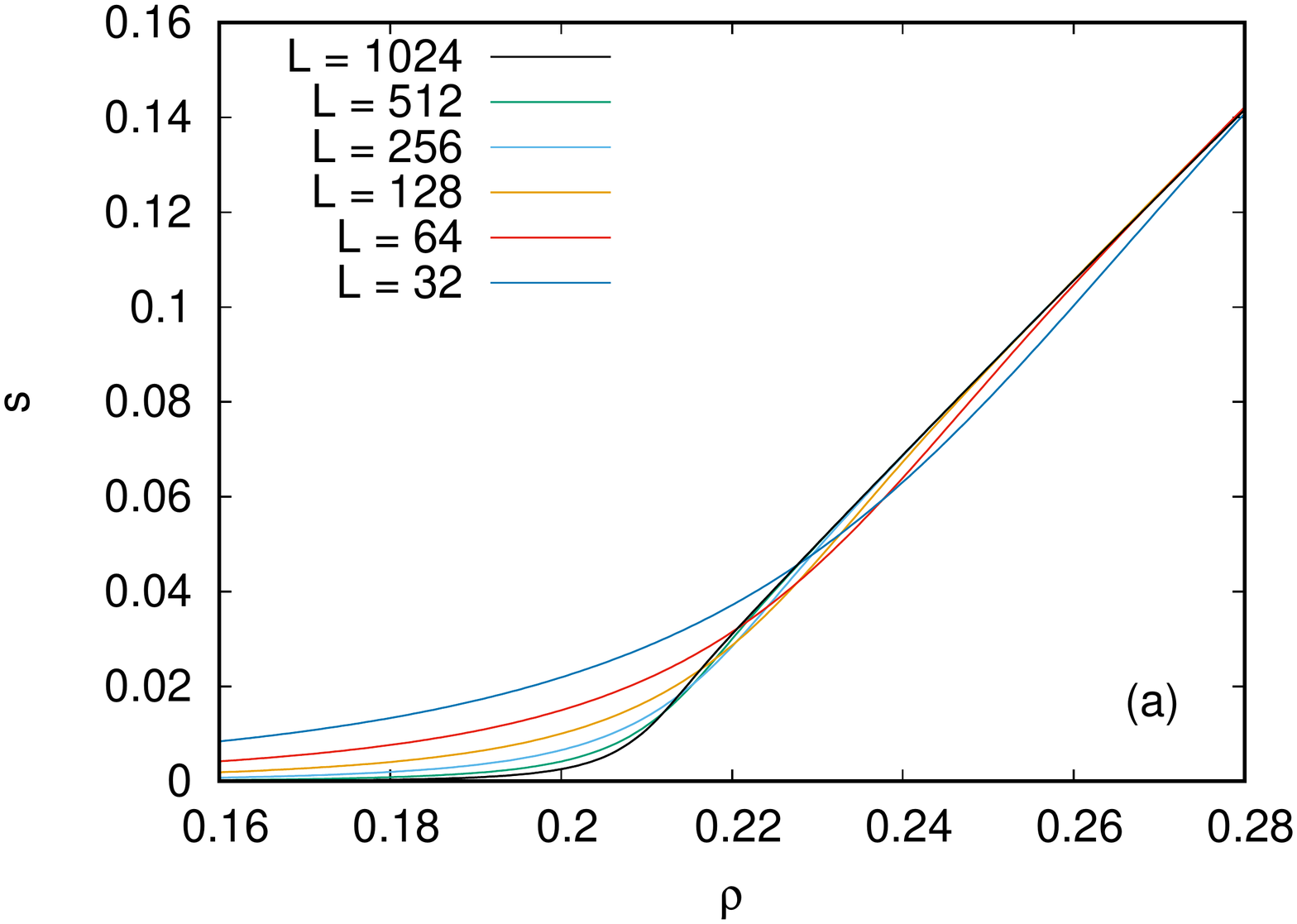,width=8.5cm, angle=0}
  \vglue -17pt
  \psfig{file=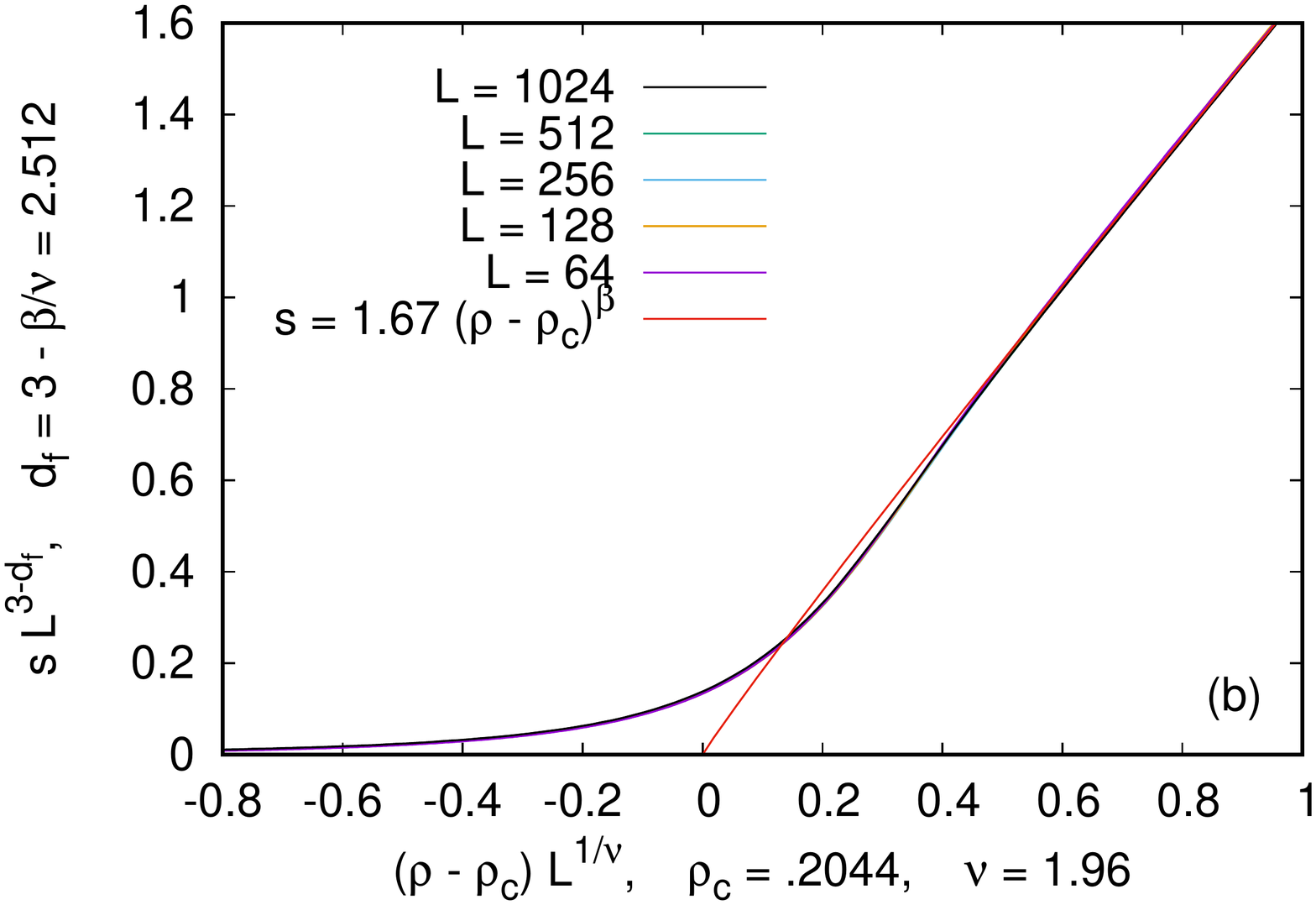,width=8.5cm, angle=0}
  \vglue -17pt
  \caption{(a) Order parameter $s$ against $\rho$ for 3D generalized Knight's move RW, for 
     different lattice sizes $L$. Notice the region very near the critical point where curves 
     cross each other (in contrast to OP and to the 2D Levy flight model discussed in the 
     previous subsection). \\
     (b) Data collapse plot of the data shown in panel a. The 
     values of $\rho_c$ and of the exponents $\nu$ and $d_f$ are fitted to obtain best collapse.
     Also plotted is a power law $s = const \, (\rho-\rho_c)^\beta$, 
     showing that Eqs.~(\ref{beta;S}) and (\ref{df-beta}) are well satisfied.}
    \label{s-3D_walks}
  \end{center}
  \vglue -11pt
\end{figure}

Here we just simulated the model with modified Knight's move RWs. As said in the 
Introduction, the finiteness of the walk trajectory does not introduce an additional length
scale in this case, whence we expect standard FSS.

Plots of the raw data of $s$ against $\rho$ for $L=64, 128, 256, 512,$ 1024, and a collapse plot 
of these data analogous to Fig.~\ref{beta-1.5} are shown in Fig.~\ref{s-3D_walks}. In contrast to 
OP and all other percolation models we are aware of, the raw data curves cross each other, but 
the scaling relations Eqs.~(\ref{beta;S}) and (\ref{df-beta}) are well satisfied. The exponent $\nu = 1.96(2)$ 
is very different from that in OP, but the fractal dimension $d_f = 2.512(10)$ is the same within
errors. These values are still preliminary (we will say more about critical exponents when 
discussing $\chi$ and gap statistics), but we can already say now that these data do not seem 
to suffer from large corrections to scaling, in contrast to those of the previous subsection.

\begin{figure}
  \begin{center}
  \psfig{file=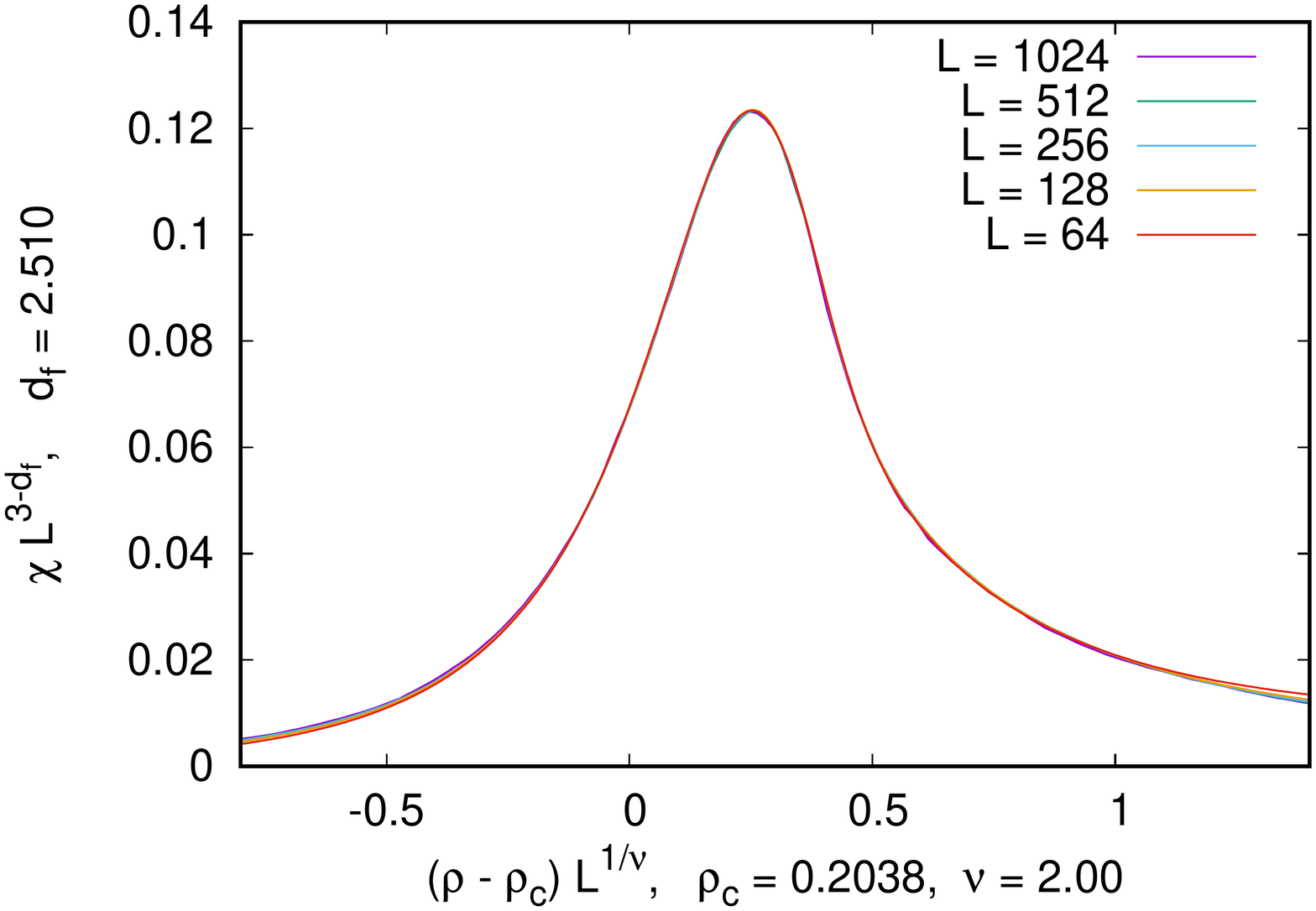,width=8.5cm, angle=0}
  \vglue -17pt
  \caption{Data collapse plot of $\chi$ against $\rho$ for 3D generalized Knight's move RWs.
     The numerical values of the critical parameters were, as in all previous collapse plots,
     obtained by eyeball fits.}
    \label{3D-chi}
  \end{center}
  \vglue -11pt
\end{figure}

A collapse plot of $\chi$ (analogous to Fig.~\ref{chi}) is shown in Fig~\ref{3D-chi}. We see 
a very good data collapse, albeit for sightly different values of the critical parameters. These
differences give a first impression of error estimates.

\begin{figure}
  \begin{center}
  \psfig{file=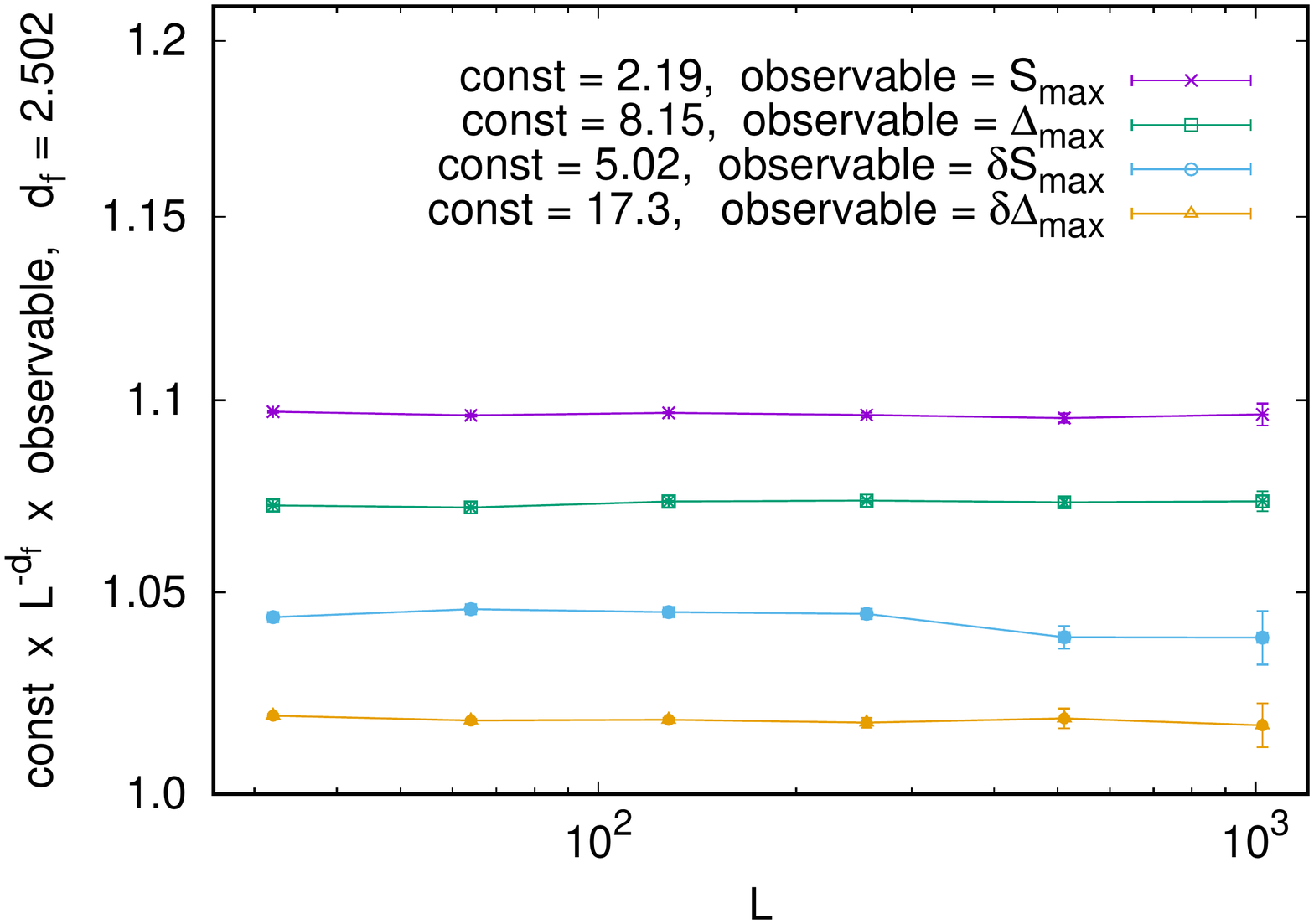,width=8.5cm, angle=0}
  \vglue -17pt
    \caption{Log-log plots, for the modified Knight's move RW in 3D, of the four event-based 
	  observables ($S_{\rm max}, D_{\rm max}$, 
    and the square roots of their variances) which should scale $\sim L^{d_f}$ at the 
    critical point. For easier comparison, each curve is shifted vertically by an arbitrary
    factor and is divided by $L^{d_f}$. Please notice the very much blown-up y scale in this
    and in the following figure.}
    \label{3D-d_f}
  \end{center}
  \vglue -11pt
\end{figure}

\begin{figure}
  \begin{center}
  \psfig{file=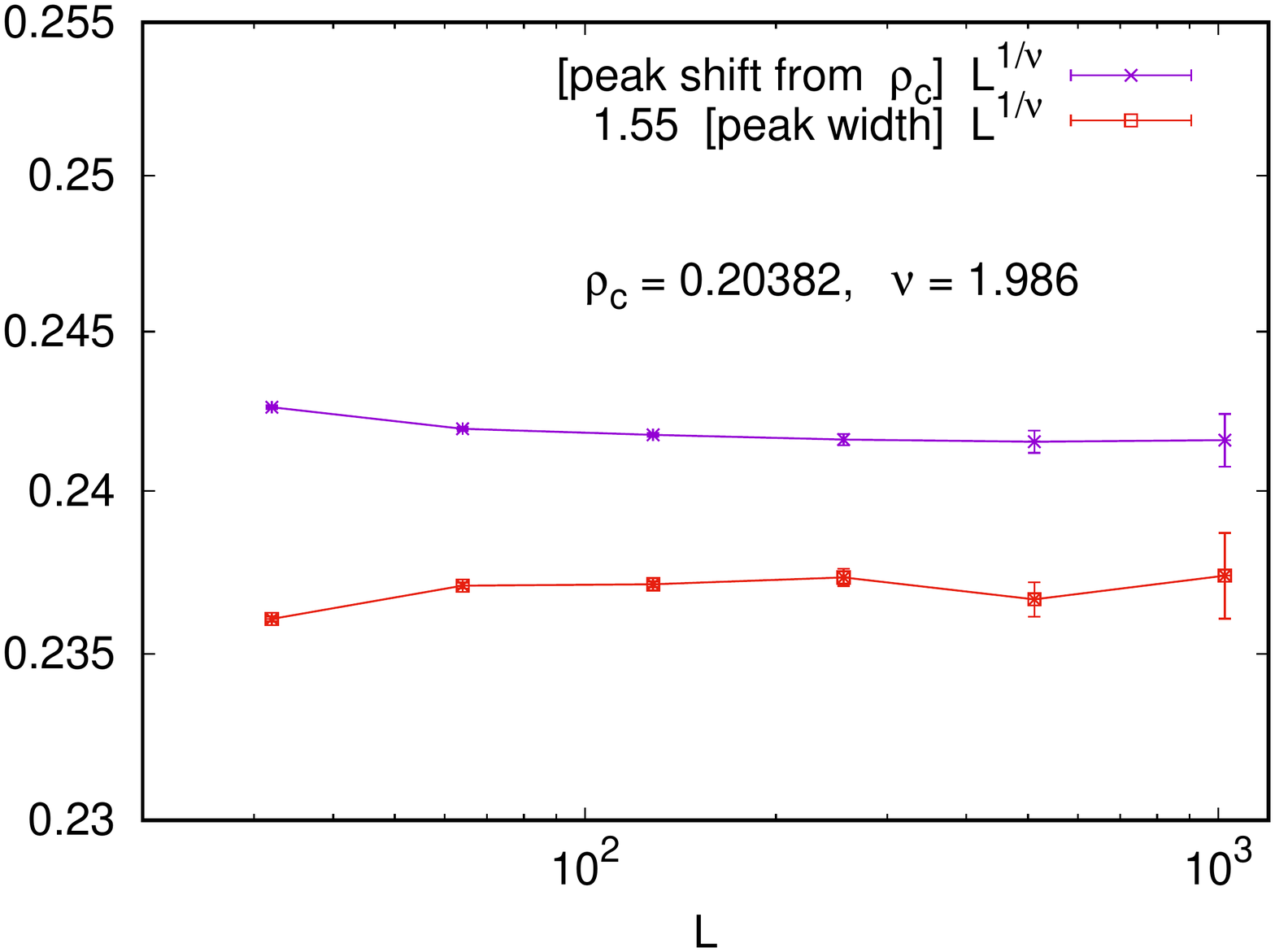,width=8.5cm, angle=0}
  \vglue -17pt
	  \caption{Log-log plots analogous to those in Fig.~\ref{nu-1.8} of the $L$-dependence of 
	  the width of the peak of $\rho_{\max}$ and
	  of its shift from $\rho_c$, but for the modified Knight's move RW in 3D. In contrast 
	  to Fig.~\ref{nu-1.8}, we find now very good scaling, with the same value of $\nu$
	  for both curves. Notice again the very much blown-up y scale.}
    \label{3D-nu}
  \end{center}
  \vglue -11pt
\end{figure}

The fact that FSS is satisfied this time with small corrections, and that critical exponents can be 
determined rather precisely, is supported by looking at event-based gap scaling. In Fig.~\ref{3D-d_f} 
we show the four observables which should scale with the fractal dimension ($S_{\rm max}, D_{\rm max}$, 
and the square roots of their variances). For easier comparison, we multiply each by an arbitrary 
constant and divide it by $L^{d_f}$. The best fit is obtained with 
\be
   d_f = 2.502(5), 
\ee
which represents our final estimate.

When determining the correlation exponent $\nu$, we are faced again with the fact that we have to
know the precise value of the critical point $\rho_c$, if we want to check that the width of the critical
peak and its shift from $\rho_c$ scale with the same power of $L$. But in contrast to the case of 
2D Levy flights, there does not seem to be a problem now, as shown in Fig.~\ref{3D-nu}.
From this figure we obtain our best estimates 
\be
    \rho_c = 0.20382(5)\;,\;\;\; \nu = 1.99(1).
\ee
It was conjectured in Refs. \cite{Weinrib-Halperin,Weinrib} that, for $3\leq d \leq 6$, one has 
$\nu = 2/a$, if the correlation decays as $C(r) \sim r^{-a}$. According to Ref. \cite{Kantor}, the 
sites visited by a RW and the sites {\it not} visited by it are correlated with $a=d-2$.
Thus the present aftermath percolation model with (generalized) Knight's move walks should be 
in the same universality class as pacman percolation in $3\leq d \leq 6$, and, in particular, 
for $d=3$ we expect $\nu=2$ in perfect agreement with our simulations \cite{Grass-pacman}. In view of this agreement,
we also conjecture that  $d_f$ and $\beta$ are simple rationals, i.e.,
\be
    \nu=2,\;\; d_f = 5/2,\;\; \beta = 1.
\ee    
This is  also compatible with the estimates of Ref. \cite{Abete}, who found $\nu=1.8(1)$ and
$\beta=1.0(1)$ for pacman percolation, and is fully confirmed by somewhat less extensive simulations
of aftermath percolation with NNN-RW's, for which $\rho_c = 0.2120(3)$.

\begin{figure}[t]
  \begin{center}
  \psfig{file=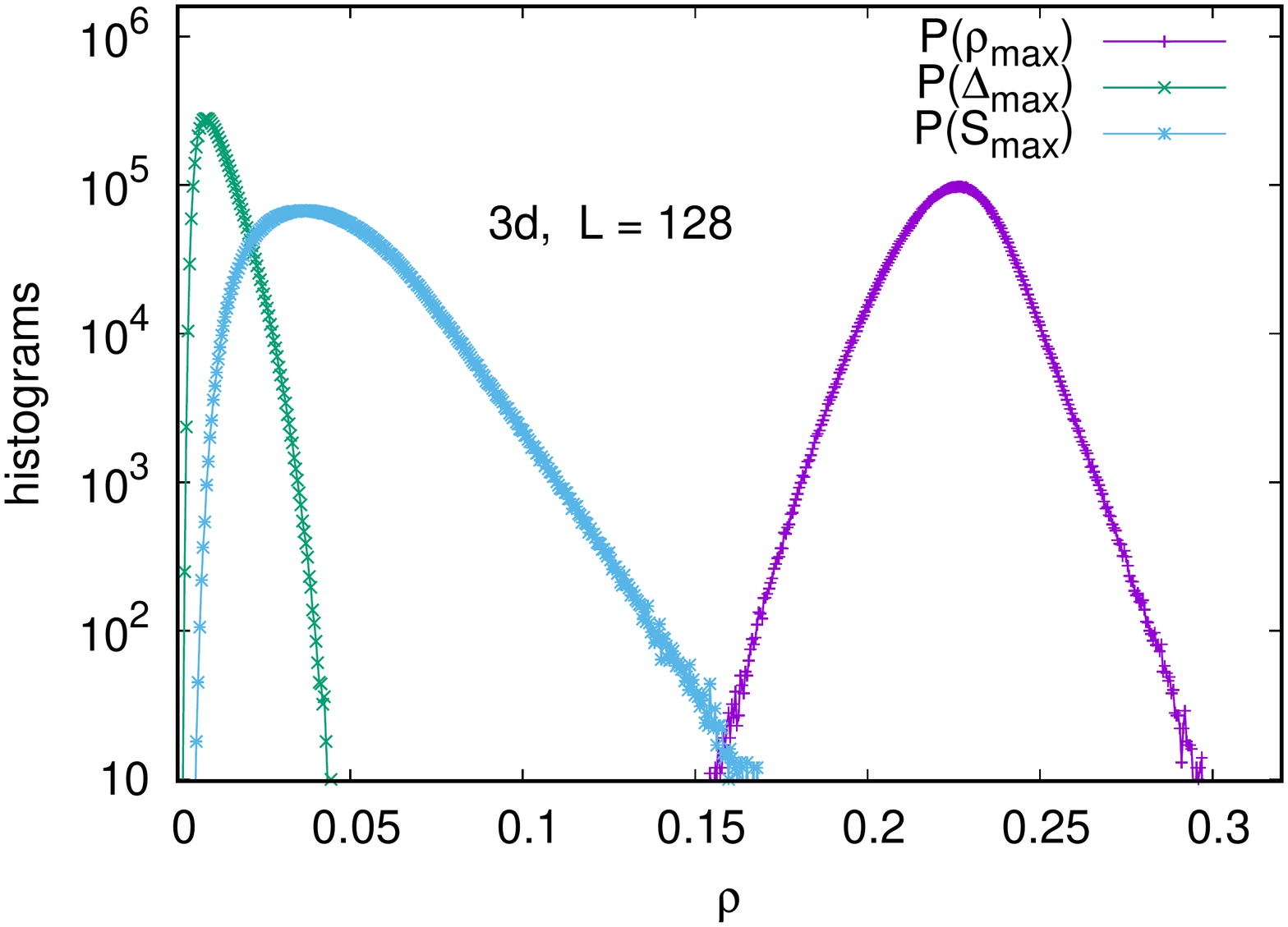,width=8.5cm, angle=0}
  \vglue -17pt
          \caption{Histograms of $S_{\rm max}, \rho_{\rm max}$, and $\Delta_{\rm max}$ for
          aftermath percolation with Knight's move RW in d=3 for $L=128$,
          based on a sample of $4\,500\,000$ realizations.}
    \label{distribs}
  \end{center}
  \vglue -11pt
\end{figure}

In the present paper we also measured the distributions of $S_{\rm max}, \rho_{\rm max}$, and
$\Delta_{\rm max}$ and their scaling functions defined in Eqs.(\ref{P_S}), (\ref{P_rho}), and (\ref{P_Delta}). It 
was claimed in Ref. \cite{Fan} that these are super-universal (i.e., universal across different
universality classes) and the same even in discontinuous percolation transitions. Due to 
the possible difficulties with scaling violations mentioned above, we postpone their 
discussions for the model with Levy flights to a forthcoming paper, where we shall also
discuss several other models. Here we present just one figure for the Knight's move RW
in three dimensions (Fig.~\ref{distribs}). In this figure we show the three distributions for 
$L=128$. According to Ref. \cite{Fan}, the distribution of $S_{\rm max}$ should be Gumbel and should
thus have an exponential right-hand tail, while the two other distributions should fall off 
faster than exponential. The opposite is true: $P_\rho(\rho_{\rm max})$ and $P_\Delta(\Delta_{\rm max})$
seem to fall off exponentially, while $P_S(S_{\rm max})$ falls off faster. More details 
will be given in Ref. \cite{Fesh-distrib}.

\section{Conclusions}

In this paper, we have introduced a new version of CP. Motivated by the 
fact that disasters like wars, floods, or hurricanes often leave a weakened region which then 
falls easy prey to a second disaster like an epidemic, we have studied percolation restricted
to the sites visited by generalized RWs. Essentially, this aftermath epidemic model
is the inverse of pacman percolation \cite{Abete,Kantor}, where percolation is restricted to 
the sites {\it not} visited by a RW.

A crucial difference from pacman percolation is that the sites not visited by ordinary RWs
are not connected, while those visited are. Thus, to obtain nontrivial percolation
in aftermath epidemics, one has to use generalized walks where the visited sites are not 
connected. We studied Levy flights in two dimensions, and Knight's move RWs both in two and three
dimensions.

In three dimensions (and with Knight's move RWs), we found that our model is in the same 
universality class as pacman percolation, and we conjecture that not only $\nu=2$ is a simple
rational, but also $d_f=5/2$.

Knight's move RWs in 2D do not lead to a sharp percolation transition. This is analogous to 
pacman percolation, where one also has to go to three or more dimensions to find a sharp
transition. But for Levy flights, sharp transitions are found whose universality classes seem 
to depend on the Levy flight exponent $\sigma$.

As a control parameter, one can take in these models the number of walker steps or the number of
visited points. Since finite walks might introduce new length scales, one has to worry that 
this breaks scale invariance and thereby violates  one of the essential assumptions in the 
theory of critical phenomena. We find that this is indeed the case for Levy flights (but not
for Knight's move RWs). Thus, it is not obvious that the usual FFS applies.
We found indeed no such problem for Knight's move RWs in 3D. But we found problems in the 
form of very poor scaling in the case of Levy flights. It is not clear whether these are 
finite-size corrections, or whether they show that FSS is basically broken in this model.
Another effect induced by additional length scales could be that different observables with
the same scaling dimension show different critical exponents. In particular, we looked 
carefully into the possibility that there are two different correlation exponents, as has been
found in some other nonstandard percolation models. We found no such deviation from FSS.

\begin{figure}[t]
  \begin{center}
  \psfig{file=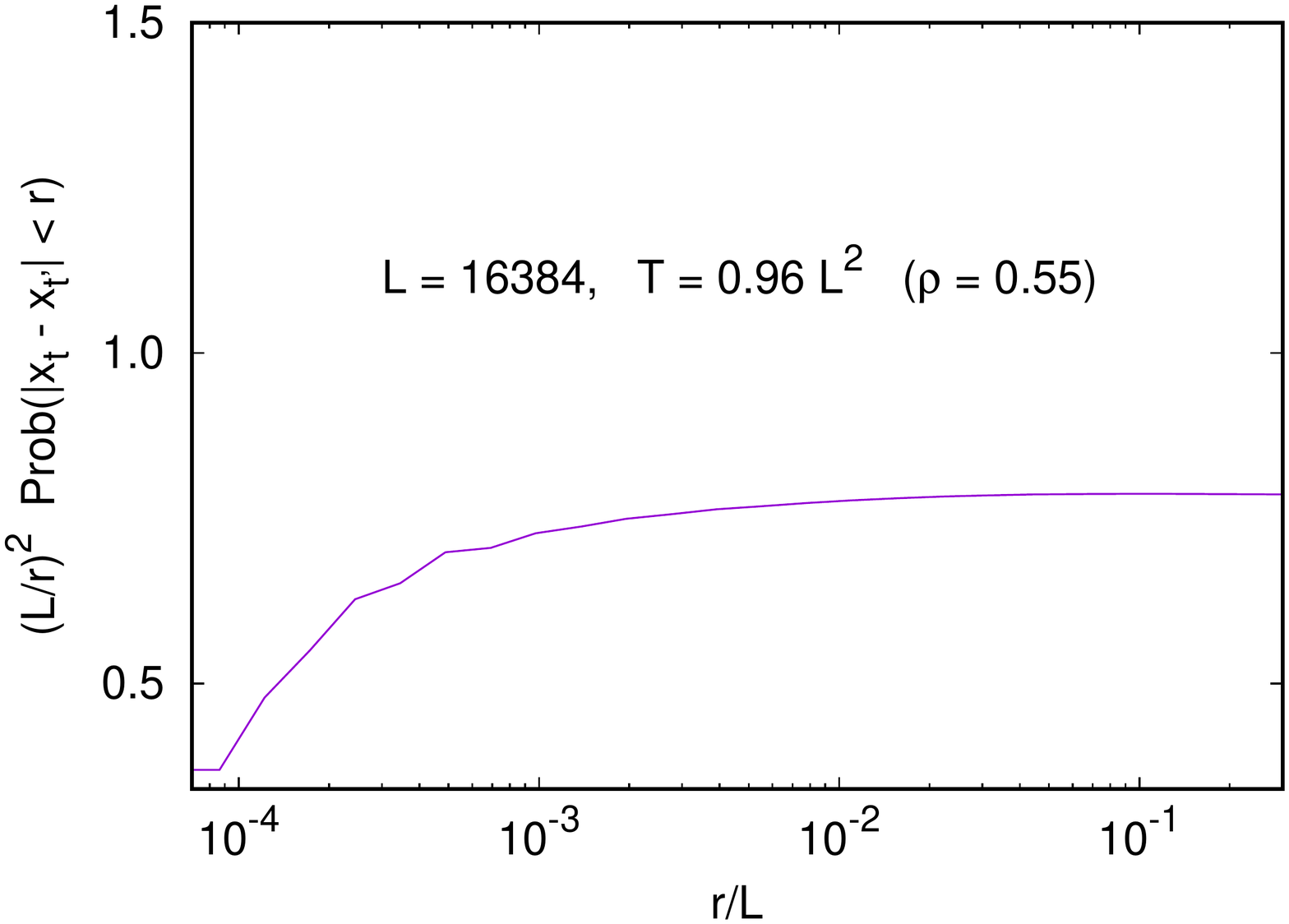,width=8.5cm, angle=0}
  \vglue -17pt
          \caption{Correlation sum of visited sites by a Levy flight $\sigma = 1.5$ with $T = 0.96 L^2$
	  steps on a square lattice of size $L = 16384$, multiplied by $(L/r)^2$ and plotted over $r/L$. 
	  The density of visited points for this $T$ is roughly
	  0.55, i.e., the critical density at the percolation threshold. The curve is obtained by averaging
	  over 2000 realizations, thus the deviations from a smooth curve are not due to noise but due to
	  the discreteness of the lattice.}
    \label{correl}
  \end{center}
  \vglue -11pt
\end{figure}

When simulating and analyzing these models, we used the fast NZ algorithm. This 
implied that we could very quickly determine quantities like cluster masses and gaps i.e.,
jumps in the leading cluster mass), but not spanning probabilities. Thus, we have not 
considered the latter, nor have we looked at backbones or conductivity exponents. But we 
have analyzed our data both within the traditional paradigm where one considers observables
at given values of the control parameter, and in the event-based ensemble \cite{Manna,Nagler,Li2023},
where observables are measured at those control parameter values where the biggest gap occurs.
We found that the latter gives, in general, more precise results.

\section{Appendix}

To measure correlations between sites visited by a Levy flight in two dimensions, we measured the 
correlation sum $C(r)$, i.e., the fraction of pairs of visited sites which are a distance $\leq r$ apart.
This is shown in Fig.~\ref{correl} for $\sigma = 1.5$, $L=16384,$ and $T=0.96 L^2$, which corresponds
to a density $\rho = 0.55$ of visited sites. For better resolution, we multiplied this by $(L/r)^2$, so 
the curve would be a horizontal flat line for a Poisson process, i.e., for $\sigma = 0$. We see 
only very small deviations from this, and definitely no power law. 

Acknowledgements: M.F. and A.A.M. acknowledge supports from the research council of the Alzahra University.
P.G. thanks Nuno Ara\'ujo, Michael Grady, Hans Herrmann, and Yacov Kantor
for discussions about CP.

\end{document}